\documentclass[%
 reprint,
nofootinbib,
 amsmath,amssymb,
 aps,
]{revtex4-2}

\usepackage{graphicx}
\usepackage{dcolumn}
\usepackage{bm}
\usepackage{rotating}
\usepackage{multirow}

\usepackage{lipsum} 
\usepackage[export]{adjustbox}

\begin{document}

\preprint{APS/123-QED}

\title{Neural-Network Decoders for Quantum Error Correction using Surface Codes:\\
A Space Exploration of the Hardware Cost-Performance Trade-Offs}

\author{Ramon W.J. Overwater}
 \email{r.w.j.overwater@tudelft.nl}
 \author{Masoud Babaie}
\author{Fabio Sebastiano}%
\email{F.Sebastiano@tudelft.nl}
\affiliation{%
 QuTech and Department of Quantum and Computer Engineering, Delft University of Technology, 2600 GA Delft, The Netherlands
}%

\date{\today}

\begin{abstract}
Quantum Error Correction (QEC) is required in quantum computers to mitigate the effect of errors on physical qubits. When adopting a QEC scheme based on surface codes, error decoding is the most computationally expensive task in the classical electronic back-end. Decoders employing neural networks (NN) are well-suited for this task but their hardware implementation has not been presented yet. This work presents a space exploration of fully-connected feed-forward NN decoders for small distance surface codes. The goal is to optimize the neural network for high decoding performance, while keeping a minimalistic hardware implementation. This is needed to meet the tight delay constraints of real-time surface code decoding. We demonstrate that hardware based NN-decoders can achieve high decoding performance comparable to other state-of-the-art decoding algorithms whilst being well below the tight delay requirements $(\approx 440\ \mathrm{ns})$ of current solid-state qubit technologies for both ASIC designs $(<30\ \mathrm{ns})$ and FPGA implementations $(<90\ \mathrm{ns})$. These results designates NN-decoders as  fitting candidates for an integrated hardware implementation in future large-scale quantum computers.
\begin{description}
\item[Data and Code]
The data is at DOI: https://doi.org/10.4121/16539786 \cite{DOI}
\end{description}
\end{abstract}

\maketitle


\section{Introduction}\label{sec:Introduction}

For certain problems, quantum-computing algorithms have been demonstrated to run with polynomial time complexity, where classical counterpart would scale with an exponential time complexity \cite{Harrow2017, Montanaro2016, Shor1997, Grover1996}.
This speed-up is ascribed to the use of quantum bits (qubits) that, unlike classical bits, can exploit quantum effects, such as superposition, entanglement and interference \cite{Nielsen2000, Feynman1982}.
Unfortunately, the information stored in the qubits can be lost via decoherence, due to their sensitivity to their environment.
The errors due to decoherence can be mitigated by adopting quantum-error-correction (QEC) schemes that encode multiple imperfect \emph{physical} qubits into a \emph{logical} quantum state, similar to classical error correction. However, while classical bits can be simply copied to introduce redundancy, the quantum no-cloning theorem prevents the copying of qubits \cite{Park1970, Wootters1982}, thus calling for ad-hoc QEC schemes.

The Surface Code (SC), a planar form of the Toric Code \cite{Kitaev1997}, is among the most popular QEC schemes thanks to its high error threshold, scalable 2D structure and the need for only next-neighbour interactions \cite{Fowler2012}. This makes it suitable for integration in promising solid-state qubit technologies, like superconducting qubits \cite{Koch2007, Schreier2008} and quantum-dot-based qubits \cite{Veldhorst2014}. Although encoding a logical state in a SC is straightforward, detecting the errors occurring on the physical qubits typically requires a complex decoder \cite{Edmonds1965, Bravyi2014, Cianci2010, Delfosse2017}, as physical qubits cannot be directly measured without losing quantum information. In addition to the computational complexity of QEC decoding algorithms, the decoder should run orders of magnitude faster than the decoherence process affecting the physical qubits, with a required execution time well below 1~$\mu$s for typical solid-state qubits. This stringent timing requirement has raised the question whether the decoders need to be implemented in hardware instead of running in software for even faster inference \cite{Varsamopoulos2018_1}. Furthermore, hardware decoders would be preferred to support the scalability of quantum computers. Promising candidates for large-scale quantum computers  comprise large arrays of cryogenic solid-state qubits controlled by local electronics also operating at cryogenic temperatures to ensure compactness and reliability by avoiding long interconnects between several temperature stages \cite{Patra2018,Sebastiano2017,Patra2020,Degenhardt2019,Lehmann2019,Voinigescu2019,Ruffino2018,Hornibrook2015,Bohuslavskyi2018}. Thus, the QEC decoder must also run at cryogenic temperature and an integrated hardware implementation is favorable to minimize the area occupation (for compactness) and the power dissipation (to comply with the limited cooling budget of cryogenic refrigerators). Recent work has demonstrated hardware-based decoders that run fast enough  \cite{holmes2020,das2020,Das2021,ueno2021}, but research is lacking on the hardware implementation of a decoding solution that has the potential to outperform them all: neural networks. 


\begin{figure*}[ht]
    \centering
    \includegraphics[width=\textwidth,center]{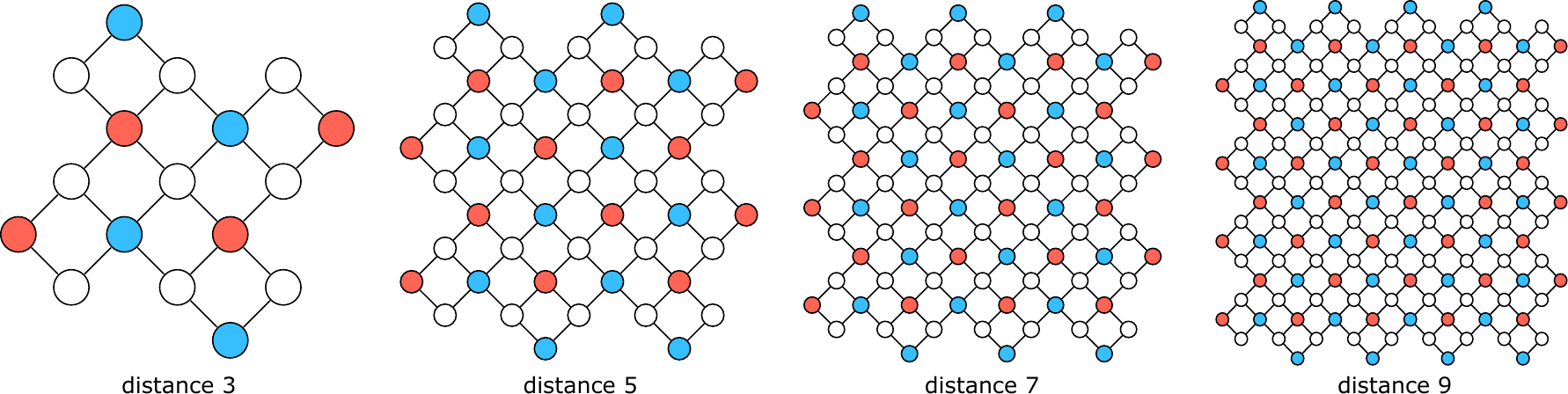}
    \caption{Schematic representation of the four smallest distances of the rotated surface code. The white dots represent the data qubits, the blue dots the $X$-ancillas and the red dots the $Z$-ancillas. The connections between the qubits correspond to the local interactions whilst performing the measurement round.}
    \label{fig:scdistances}
\end{figure*}

Neural network (NN) decoders have attracted large interest, thanks to their fast and constant inference time and state-of-the-art decoding performance \cite{Varsamopoulos2018_1, Varsamopoulos2018_2, Fosel2018, Baireuther2017, Torlai2017, Krastanov2017, Chamberland2018, Ni2018, Varsamopoulos2019}. The hardware requirements for neural network decoders have been estimated before \cite{Varsamopoulos2018_1, Chamberland2018}, but the trade-offs between hardware cost and performance have not been explored. This work bridges this gap by focusing on the hardware implementation of a neural-network decoder for SC QEC \cite{Varsamopoulos2018_2}.
First, the relation between decoder performance and the NN design parameters, such as the number of layers and their size, the neuron transfer function, signal quantization, and symmetries, are explored. Then, the trade-offs between the decoding performance (error rate, computing delay) and the hardware cost (area, power) are evaluated for an implementation on both an application-specific integrated circuit (ASIC) and a commercial FPGA, with explicit attention to cryogenic operation of both platforms.

This work demonstrates that hardware NN-based decoders can achieve high decoding performance comparable to other state-of-the-art decoding algorithms while satisfying with ample margin the tight delay requirements of current solid-state qubit technologies. The hardware cost in terms of silicon area and power quickly increases with neural network size and surface code distance. The obtained decoding times are low enough for future work to explore further optimization of the hardware costs.

The rest of this paper is organized as follows. First, section \ref{sec:SurfaceCodes} gives a short background on decoding the surface code. This is followed by section \ref{sec:Decoder}, which shows the proposed decoder. Next, section \ref{sec:Setup} outlines the simulation setup. The decoding performance results and design space exploration are shown in section \ref{sec:Decoding_Results}. The results are combined with the hardware cost estimations in section \ref{sec:Hardware_Results}. The results are then discussed in section \ref{sec:Discussion}, and finally, conclusions are drawn in section \ref{sec:Conclusion}.

\section{Decoding the Surface Code}\label{sec:SurfaceCodes} 
The surface code, as shown in Fig. \ref{fig:scdistances}, is a simple 2D scalable structure of physical qubits (denoted by the dots) that only requires local interactions between qubits (as illustrated by the lines in between the dots). Only a brief overview of the surface-code operation is given in this section; the interested reader is referred to \cite{Fowler2012} for a complete treatment. This work focuses on \emph{rotated} surface codes \cite{Horsman2012}, which use the least amount of physical qubits per logical qubit. Each code has a distance $d$, meaning that a perfect decoder can correctly identify a maximum of $(d-1)/2$ physical errors. A rotated surface code of distance $d$ consists of a $d\times d$ grid of data qubits (white dots in Fig. \ref{fig:scdistances}) that encode a single logical qubit. The $d^2 -1$ colored dots in Fig. \ref{fig:scdistances} represent two types of ancilla qubits. These $X$ and $Z$-ancillas can be measured to find the errors on the adjacent data qubits without destroying the quantum state of the encoded logical qubit. This measurement outcome is called the error syndrome and needs to be continuously measured to detect errors in every so called surface code cycle. The task of the decoder is to find the errors on the data qubits from this ancilla measurement syndrome.

\subsection{Surface Code Cycle}
During a SC cycle, all ancillas are first initialized into the ground state $|0\rangle$. Next, the $X$-ancillas ($Z$-ancillas) are brought onto the $x$-axis ($z$-axis) using a Hadamard gate (identity gate\footnote{The identity gate (idling) is shown to keep the two execution flows synchronized.}).

\begin{figure}[b!]
  \centering
  \includegraphics[width=0.3\textwidth,center]{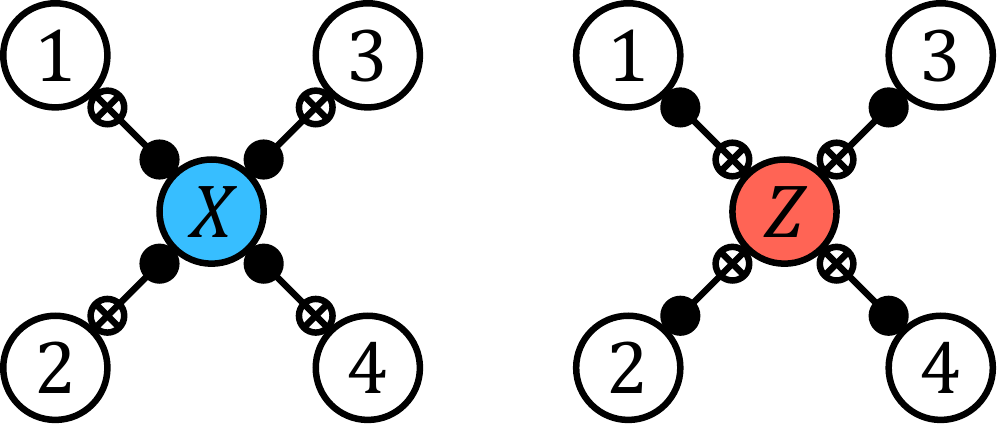}
  \caption{CNOT gate sequence. The number on each data qubit indicates the order in which the four adjacent data qubits are addressed by both the $X$-ancilla (blue, on the left) and the $Z$-ancilla (red, on the right). This process is performed on each ancilla qubit of the rotated surface code shown in Fig. \ref{fig:scdistances}.}
  \label{fig:cnotdance}
\end{figure}

\begin{figure}[b!]
  \centering
  \includegraphics[width=0.5\textwidth,center]{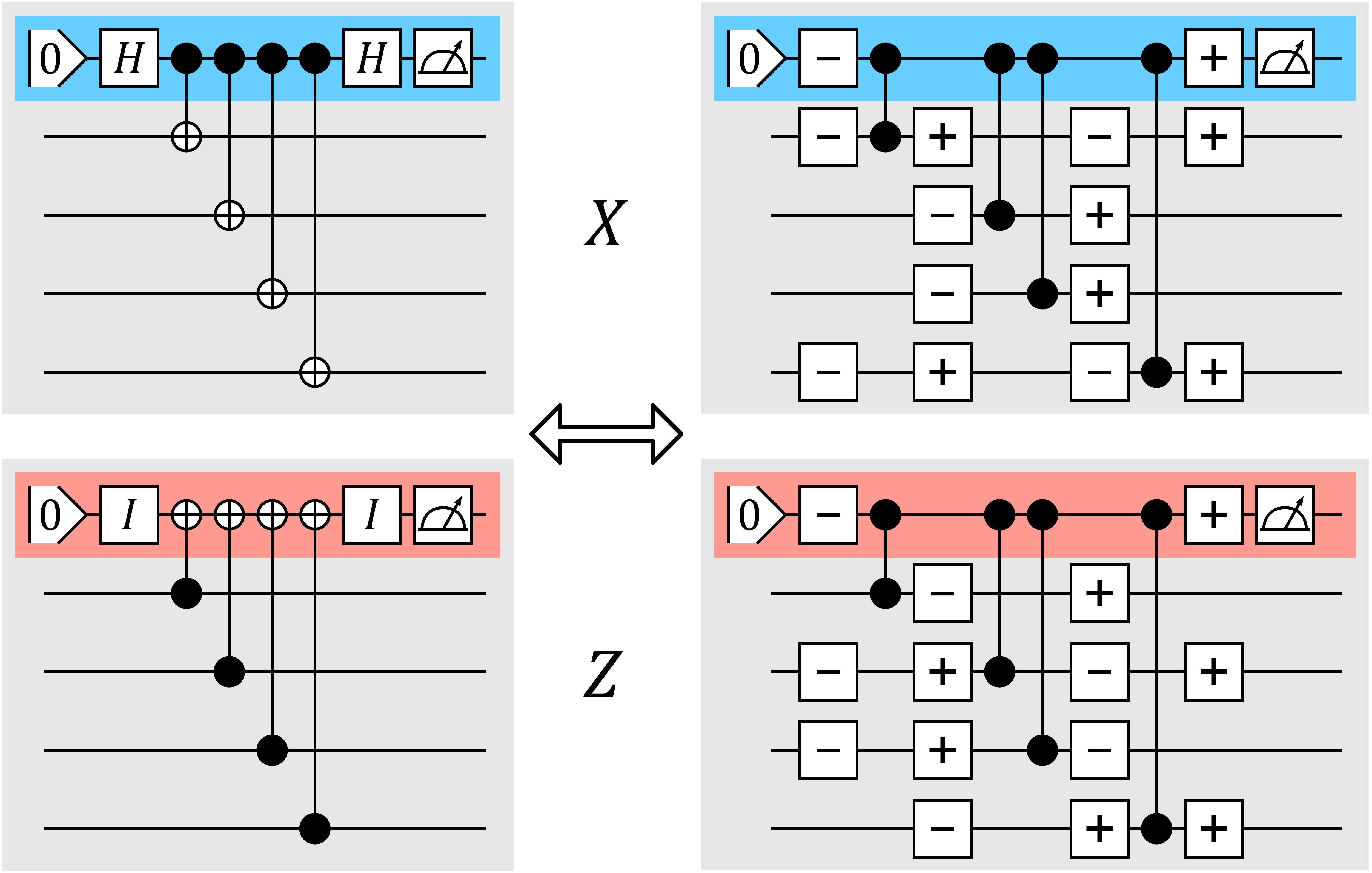}
  \caption{(left) Quantum circuits for the commonly used surface code cycle employing Hadamard and CNOT gates. (right) Equivalent circuit with $CZ$ and $R_y\left(\pm\frac{\pi}{2}\right)$ on the right. The $R_y\left(\pm\frac{\pi}{2}\right)$ are denoted as $\pm$. In both figures the top circuit shows the circuits for the $X$-ancillas and the bottom circuit for the $Z$-ancillas. The colored qubits are the ancillas. The other gray qubits are the data qubits surrounding this ancilla. Top to bottom this is the same order as the 1 to 4 shown in Fig. \ref{fig:cnotdance} in the CNOT dance.}
  \label{fig:sccycle}
\end{figure}

A sequence of CNOT gates is then performed, as shown in Fig. \ref{fig:cnotdance}, to entangle each ancilla  with its four adjacent data qubits. In case an ancilla is at the edge of the SC, only the two neighbouring data qubits are used. Finally, the $X$-ancillas are brought back onto the $z$-axis and all ancillas are measured in the $z$-basis. This whole cycle is shown in Fig. \ref{fig:sccycle} on the left. \\

The measurement on each ancilla will return either +1 or -1, reflecting the parity of the four (or two) adjacent data qubits. The set of all ancilla measurements in a SC is called the \emph{syndrome}. After all the ancillas are pushed into a state, there are still some degrees of freedom of the surface code. These degrees of freedom define the logical state of the logical qubit. Any operation that does not change the syndrome and logical state is called a \emph{stabilizer}. The next section shows this in more detail. Measuring the ancillas puts the total surface code into an eigenstate of all stabilizers where the syndrome represents the eigenvalues. 

After the first measurement cycle, the surface code is initialized and all ancillas will either be +1 or -1. These do not represent errors, but the random initial quiescent state of the data qubits. Repeated measurement cycles will keep it in the same quiescent state. Any change in the syndrome after measurement indicates a deviation from the quiescent state and thus an error.

\subsection{Logical Operations and Errors}
\begin{figure} [b!]
    \centering
    \includegraphics[width=0.4\textwidth,center]{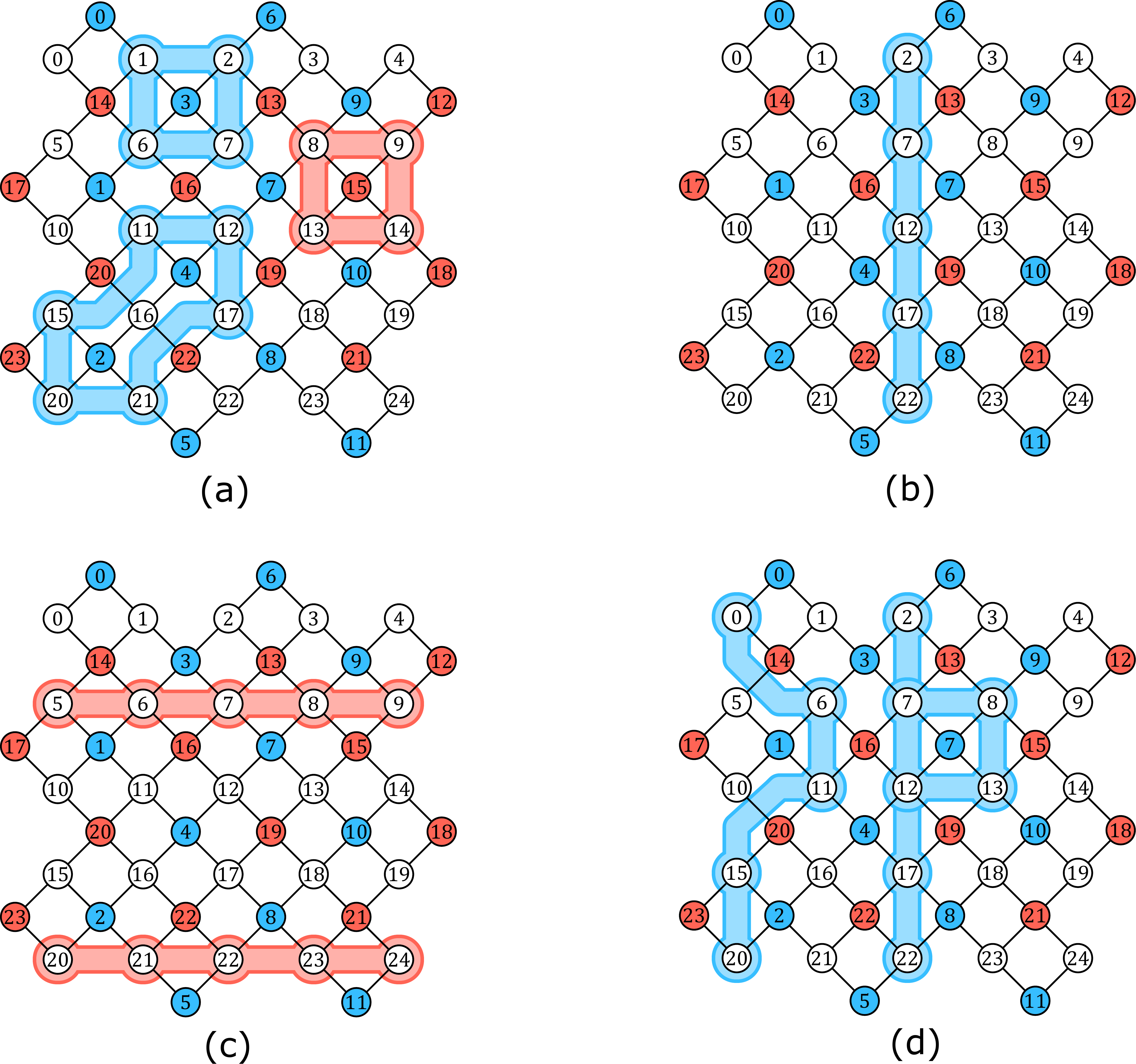}
    \caption{(a) The product of four stabilizers on the distance 5 surface code. (b) A single logical $X$ operations. (c) Two logical $Z$ operations. (d) Two illustrations of the product of a logical $X$ operation and an $X$ stabilizer.}
    \label{fig:logicaloperators}
\end{figure}

To understand how a stabilizer does not change the syndrome, see Fig. \ref{fig:logicaloperators}(a). Performing four $X$-operations (shown in blue) on the data qubits around $X$-ancilla 3, do not change the parity of the (13, 14, 16) $Z$-ancilla measurements and thus does not change the syndrome. A similar reasoning applies to the $Z$-operations around $Z$-ancilla 15.
A product of two stabilizers is performed around $X$-ancillas 2 and 4. Due to the double $X$-operation on data qubit 16, an identity operation is performed on that qubit. It can be seen that this product of two stabilizers also does not change any of the adjacent $Z$-stabilizer measurements and also forms a continuous loop of single qubit operations.

In general, a product of $X$ or $Z$-stabilizers will always form a closed chain (or loop) of $X$ or $Z$ single-qubit operations. Thus, these loops will never change the quiescent state and the syndrome. Since an error is modeled here as a random and non-intentional operation on a data qubit, errors that, by chance, form loops will not change the syndrome measurement and the logical state.

Next, Fig. \ref{fig:logicaloperators}(b) shows a chain of $X$-operations running between the top and bottom edge. Such an edge-to-edge chain performs a logical $X$ operation. One can check that this does not affect any of the $Z$-ancilla parity measurements.

Similarly, Fig. \ref{fig:logicaloperators}(c) shows two logical $Z$-operations, which should correspond to a logical $I$-operation. Not altering the logical state and syndrome, those operations could also be written as a product of stabilizers. In general, any even number of logical operations can be written as a product of stabilizers and every odd number of logical operations can be written as a single logical operation and a product of stabilizers.

Finally, Fig. \ref{fig:logicaloperators}(d) shows a product of a logical $X$-operation as a chain between data qubit 2 and 22 with a stabilizer around $X$-ancilla 7. A product of these two results in a chain with the same shape as the one drawn between data qubit 0 and 20 and with the same effect on the logic state. This exemplifies that any odd number of chains, not necessarily straight, between the top and bottom will result in a logical $X$-operation. Similarly, chains between the left and right sides of the surface code results in logical $Z$-operations.

\subsection{Pure Errors}
\begin{figure} 
    \centering
    \includegraphics[width=0.5\textwidth,center]{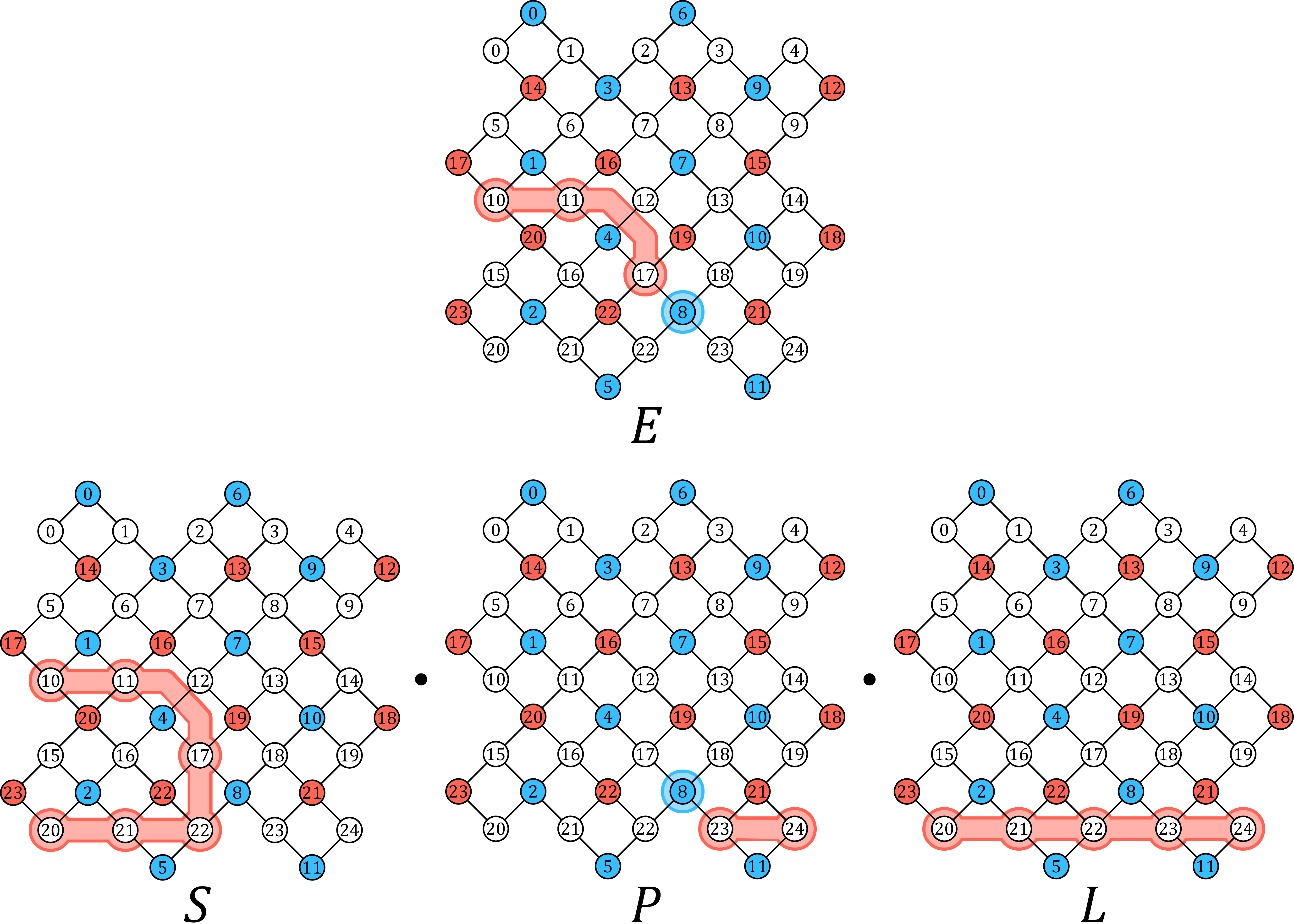}
    \caption{The decomposition of the error $E$ into a product of stabilizers $S$, a logical error $L$ and a pure error $P$. Of this product, only the pure error leads to a change in the syndrome outcome.}
    \label{fig:error_decomposition}
\end{figure}

The errors and operators discussed in the previous section do not change the measurement syndrome. Note that these operation chains do not end in the center of the surface code. If they do end in the center, the measurement syndrome will change. For instance, the $Z$-error chain shown in Fig. \ref{fig:error_decomposition} denoted by the error $E$ starts at the edge at data qubit $10$ and ends in the center at data qubit $17$. As this is not a product of just stabilizers and logical operators, errors on the measurement syndrome will be triggered, in this case at $X$-ancilla 8.

As Fig. \ref{fig:error_decomposition} shows, any stabilizer $S$ or logical operators $L$ can be applied on top of the error $E$ without changing the error syndrome. We call any state that gives the same error syndrome as the original error, and thus is only separated by stabilizers and logical operators, a pure error $P$ \cite{Varsamopoulos2018_1, Poulin2006}. This pure error can thus be logically different from the original error $E$, but it will always give the same error syndrome. To phrase differently, any error can be decomposed in a product of stabilizers, a logical operator and a pure error. 

\subsection{Decoding}
The purpose of the decoder is to identify an error configuration on the data qubits that produces the same syndrome as was measured and is logically equivalent to the actual data error configuration. In other words, the decoder must output any data error configuration that only differs from the actual data error configuration by a product of stabilizers. This can then either be used to immediately correct the errors or can be tracked for later correction using Pauli frames \cite{Riesebos2017}. The error syndrome must be the same to ensure that the surface code returns to a logical state. The logical error must also be the same to prevent logical errors during the computation. As stabilizers do not influence either the syndrome or the logical state, they can be neglected.

Since many data qubit configurations produce the same syndrome, the error syndrome generation is a non-invertible function, thus making it impossible to unambiguously find the real data qubit configuration. This constitutes the main challenge for the decoder implementation and necessarily requires the use of heuristics. Splitting the task into a separate pure error decoder and a logical error decoder could ease this task.

\subsection{Decoding Performance}
\begin{figure}[b]
        \centering
        \includegraphics[width=0.45\textwidth,center]{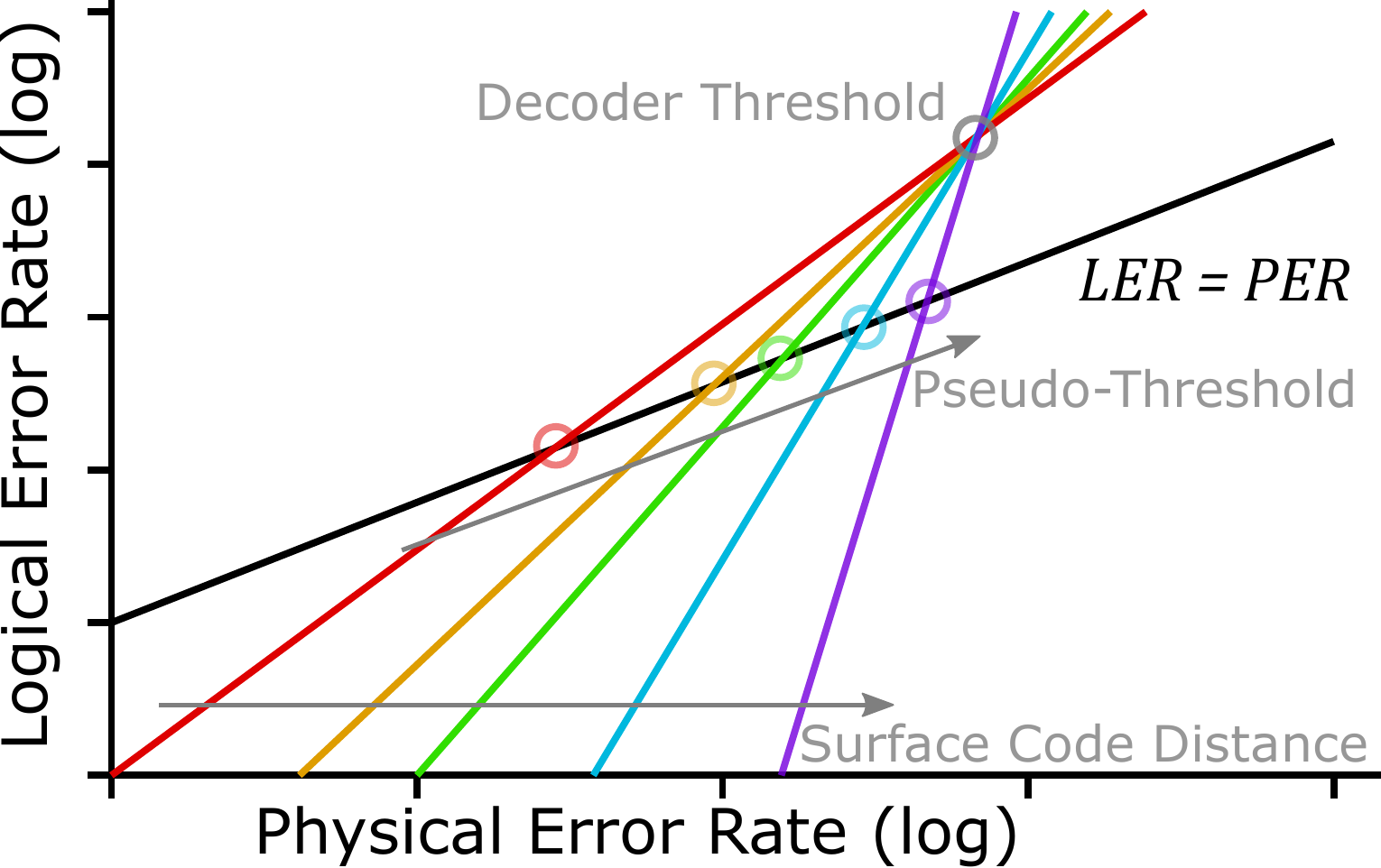}
        \caption{Sketch of the logical versus physical error rate of an un-encoded qubit (black line) and five different distances of the surface code. The pseudo-threshold is shown by colored circles and the decoder threshold by the gray circle. For a larger code distance, both the slope and the pseudo-threshold increase.}
        \label{fig:threshold}
    \end{figure}

    In order to quantify the decoding performance, the relation between the  error rate of the logical qubit and the error rate of the physical qubits must be analyzed, as shown in  Fig. \ref{fig:threshold}. In this figure, both the logical error rate for an increasing surface-code distance (colored lines) and the error rate for an un-encoded physical qubit (black  line) are shown. The physical error rate for which the decoder achieves approximately\footnote{In practice, it is possible that all lines do not cross exactly in a single point} the same performance independent from the SC distance is defined as the \emph{decoder threshold}. 
        
    For any physical error rate below the decoder threshold, it pays off to invest in a larger distance. The decoder threshold is often used as a single parameter to quantify the performance of a decoding algorithm. However, as shown in this sketch, operating at this physical error rate will be outperformed by a single un-encoded qubit. 
    
    The physical error rate at which the logical qubit will outperform the physical qubit is called the \emph{pseudo-threshold} ($p_{th}$). The $p_{th}$ is different for every distance and is used to compare decoders at the same surface code distance. A higher $p_{th}$ is preferred as it allows obtaining an advantage of using QEC with worse qubits. Even for a fixed physical error rate well below the pseudo-threshold, a higher pseudo-threshold will still give a lower logical error rate, assuming a constant slope of the lines in Fig. \ref{fig:threshold}. 
    Thus, the decoder slope is also an important parameter. Both the slope and the pseudo-threshold increase when going to larger distances, but due to the exponential relation, the slope typically dominates the decoding performance at lower physical error rates.
    
    As this paper mainly compares decoders operating at a certain surface code distance, we will focus on comparing the pseudo-threshold and the decoder slope. The proposed decoders will be benchmarked against the Minimum Weight Perfect Matching (MWPM) algorithm \cite{Edmonds1965}, also known as Blossom or Edmonds algorithm, as it is a common algorithm used for decoding.
    
\begin{figure}
        \centering
        \includegraphics[clip,trim=0cm 0cm 0cm 0cm, width=0.55\textwidth,center]{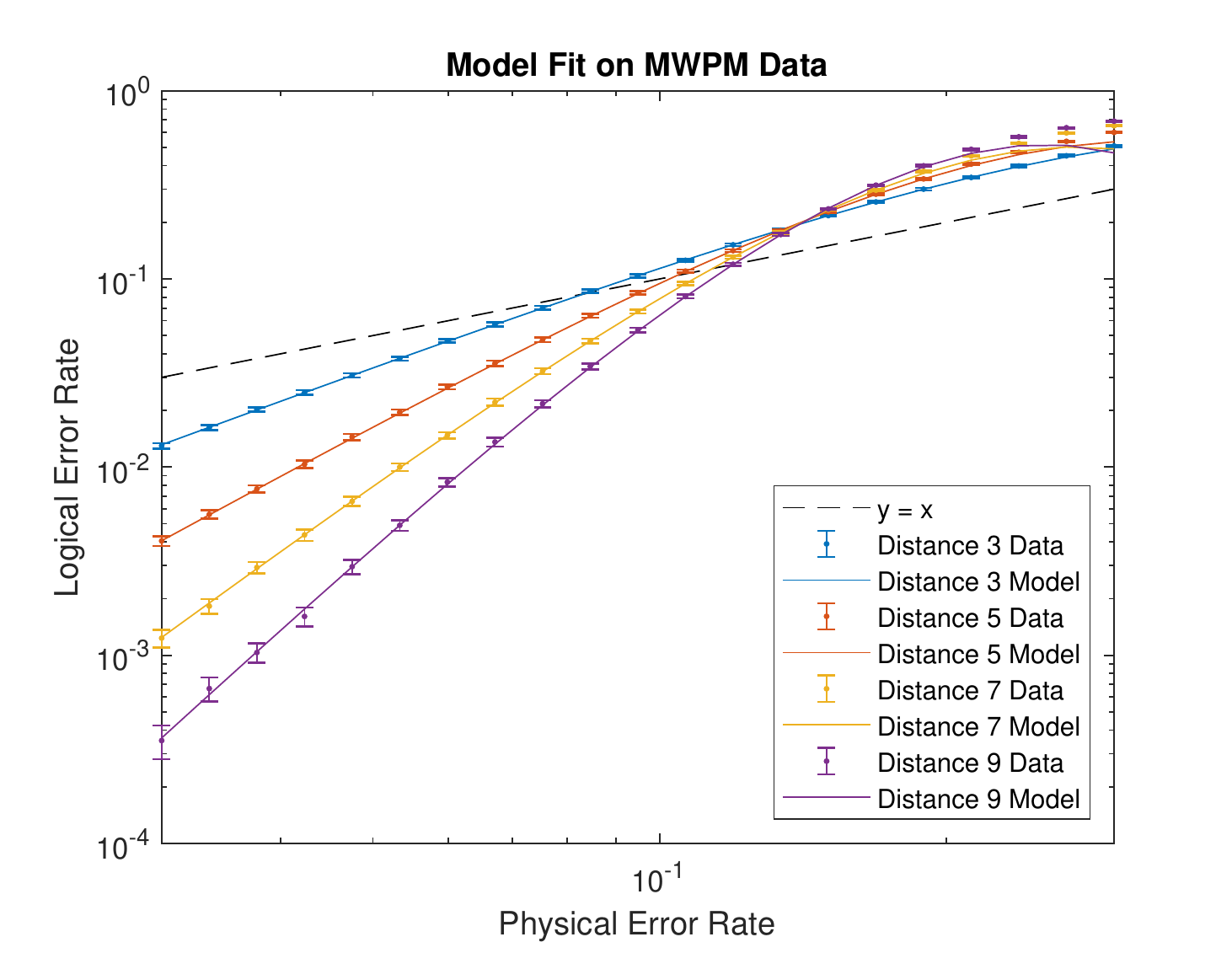}
        \caption{Model fit of (\ref{eq:model}) on the simulation of the MWPM decoder for the distance 3, 5, 7, and 9. The error bars represent the 99.9\% confidence interval.}
        \label{fig:blossomslope}
    \end{figure}

    As an example, Fig. \ref{fig:blossomslope} shows the simulated performance of the MWPM decoder for the smallest four surface code distances. The data is fitted using equation \ref{eq:model}, adapted from \cite[eq.~11]{Fowler2012} 
    where $\epsilon_p$ and $\epsilon_l$ are respectively the physical and logical error rate, and $p_{th}$, $s$  and $c$ are fitting parameters representing the pseudo-threshold, the slope for $\epsilon_p \ll p_{th}$ and the flattening of the curve for increasing physical error rates, respectively. The good fitting of the model up to the decoder threshold in Fig. \ref{fig:blossomslope} indicates that interpolation in the logarithmic domain is necessary to calculate the pseudo-threshold.
    
        \begin{equation}\label{eq:model}
        \epsilon_l = p_{th} \left(\frac{\epsilon_p}{p_{th}}\right)^{s\cdot(1 - c\cdot \epsilon_p)}
    \end{equation}
    
\subsection{Hardware Requirements and Costs}
In addition to the decoding performance, decoder implementations must also be compared based on their hardware requirements (delay) and hardware costs (area and power).

When using quantum error detection with Pauli frames, the main requirement  is  the minimum decoder throughput to avoid a data backlog \cite{Fowler2012,Riesebos2017}. In principle, the decoder can run in parallel with the main algorithm execution and, to ensure the throughput, the decoding delay should be just lower than the measurement cycle, but not necessarily much smaller. However, tracking of errors is not enough when using non-Clifford gates, and the physical  correction of errors is needed \cite{Riesebos2017}. Since such a correction must be performed before the next cycle after the error detection, the decoding can only take a fraction of the cycle time. 

The maximum allowed delay in case of transmons and silicon-based single-electron spin qubits are estimated in Table \ref{tab:sc_cycle_times} assuming the surface-code cycle shown in Fig. \ref{fig:sccycle}. For the targeted qubit technologies, the circuit on the left of Fig. \ref{fig:sccycle} can be replaced with the circuit on the right \cite{Versluis2017, OBrien2017}, as the $cz$ and $R_y(\pm\pi/2)$ gates can be performed faster and more accurately in those physical platforms.
The minimum reported duration for each operation is used to obtain the most stringent constraint on the delay. Also, the initialization of the ancillas is neglected \cite{OBrien2017, Versluis2017} to get a lower bound of the estimated cycle time. For quantum error detection, the delay must be below 440 ns. However, when including a correction step for using non-Clifford gates, the delay needs to be as small as possible. This work will thus strive to minimize the delay and report the corresponding power and area.

Since the decoder is used once per QEC cycle and the cycle duration is fixed by the qubit technology, the hardware cost in terms of power is accounted for by computing the energy per decoding cycle. The dissipated energy per cycle should be as small as possible to allow for the largest number of logical qubits before running into the cooling power limitations. When fully integrating the decoder with the qubits on the same chip (or in the same package), the area must also  be as small as possible to ease the integration requirements \cite{boter2021}.\\

\begin{table}
 \centering
 \caption{State-of-the-art operation times for the two target technologies, and the resulting surface-code duration assuming the circuit in Fig. \ref{fig:sccycle}(right). }
\begin{tabular}{c c c}
        \hline
        \hline
   Operation & Transmons & Single-Electron Spin\\
            &           & Qubit (Silicon) \\
  \hline \rule{0pt}{1\normalbaselineskip}
  Single Qubit Gate & 20 ns \cite{vanDijk2018_2} & 1 $\mu$s \cite{vanDijk2018_2} \\
  Two Qubit Gate & 40 ns \cite{vanDijk2018_2} & 0.1 $\mu$s \cite{vanDijk2018_2} \\
  Measurement & 200 ns \cite{Jeffrey2014} & 1 $\mu$s \cite{Barthel2010} \\
  Surface-code cycle duration & 440 ns & 5.4 $\mu$s\\
  
        \hline
        \hline
    \end{tabular}
 \label{tab:sc_cycle_times}
\end{table}

\section{Proposed Decoder}\label{sec:Decoder} 
In this work, we focus on decoding the smallest four rotated surface codes, see Fig. \ref{fig:scdistances}. The goal is to obtain a decoder that has high decoding performance, runs fast enough to avoid a data backlog and can be efficiently implemented in hardware.

Neural networks are a promising solution for several reasons. First, they have shown higher pseudo-threshold compared to other decoders, such as the MWPM algorithm \cite{Varsamopoulos2018_1}. They can also adapt to many error models during training, perhaps even tailored to a specific qubit technology or even an individual quantum computing sample. After training, their execution (inference) time is constant and independent of the input. On one hand, the inference time of NN decoders in hardware implementations has been estimated before \cite{Varsamopoulos2018_1, Chamberland2018}, but did not satisfy the throughput requirement.  On the other hand, these analyses do suggest that an optimized design on an ASIC could meet such a requirement. Finally, their regular structure makes them well suited for hardware optimization by parallelization and pipelining.

\begin{figure} 
    \centering
    \includegraphics[width=0.5\textwidth,center]{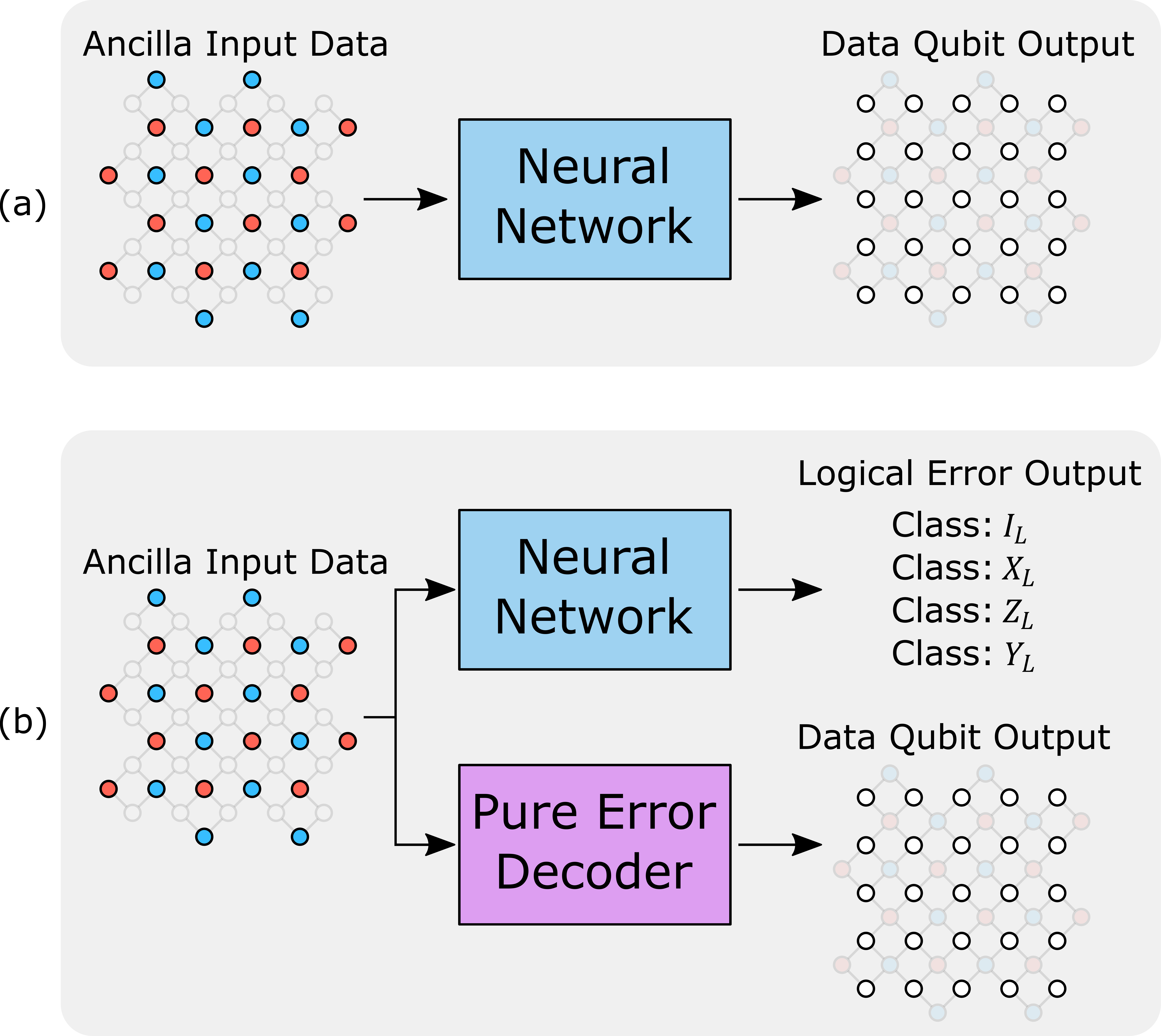}
    \caption{(a) A neural network in a low level decoder. This takes the syndrome as inputs and gives the data qubit errors as output. (b) A neural network together with a pure error decoder in a high level decoder. Here the pure error decoder gives the data qubit errors. The neural network outputs the expected logical error that the pure error decoder makes compared to the actual data qubit errors.}
    \label{fig:outputsped}
\end{figure}

However, they also have their drawbacks. A large enough training data set is needed to avoid over-fitting. Even though any data set can be generated using an error model, the size requirement of this dataset can still be a problem \cite{Varsamopoulos2018_2}. Next, neural networks are quite complex and self-trained algorithms that are difficult to thoroughly understand, thus risking to unexpectedly fail  in untested situations. On top of that, there are a lot of additional parameters that need to be optimized during training \cite{Chamberland2018, Varsamopoulos2018_2}, making the search space for finding the optimum solution even greater.
The main challenge, however, is that NN are not well suited for direct application to the decoding problem. Fig. \ref{fig:outputsped}(a) shows the NN in such a direct, so-called Low Level Decoder (LLD) application. In this configuration the NN takes in the error syndrome and guesses the error on every data qubit. The goal is that this data qubit configuration returns the correct logical error and also results in the same error syndrome as was measured. The problem is that the neural network has no notion of what such a valid solution entails. This will limit the chance that a valid data error configuration is obtained. Consequently, a re-run of the algorithm is needed until a satisfying solution is found.
To circumvent this limitation, we adopt the solution proposed in  \cite{Varsamopoulos2018_1} and use a High Level Decoder (HLD). In a HLD, the task of obtaining any correct error syndrome is performed by a Pure Error Decoder (PED). This reduces the task of the neural network to finding the type of logical error, allowing the neural network to be a classifier, a task that is well suited to neural networks. See Fig. \ref{fig:outputsped}(b). 

Other drawbacks are only relevant as the surface code distance increases. Thus, for near term small-distance surface code, the hardware and performance advantages outweigh the disadvantages. The next subsections will explain the basic functionality of the neural network and the pure error decoder chosen in this work.

\subsection{Pure Error Decoder}
The only task of the pure error decoder (PED) is finding a configuration for the data qubit errors that produces the error syndrome measured by the ancillas \cite{Varsamopoulos2018_1, Poulin2006}. As shown in Fig. \ref{fig:error_decomposition}, this pure error will only differ from the actual data qubit errors by a product of stabilizers and logical operators.
As the product of stabilizers does not influence the syndrome or the logical error, it can be neglected. Thus, the only significant difference between the error estimation given by the pure-error decoder and the effective error is a logical error. Leaving the task of the neural network to guess the logical error.

The benefit of this approach is that the guess made by the PED does not need to be the most probable. As a result, the PED can be optimized for other properties, and in this work we focus on three main points:
\begin{itemize}
    \item \textbf{Software simulation speed}. As the pure error decoder must run every time the neural network is run or trained, the speed of the pure error decoder must be maximized.
    \item \textbf{Hardware simplicity}. The area and power of the PED must be minimized.
    \item \textbf{Exploiting symmetries}. The surface code is characterized by several symmetries that can be exploited in the training of the neural network. However, as the neural network also learns on the basis of the PED output, the PED should also show the same symmetries as the surface code for fully optimizing the NN training.
\end{itemize}

\begin{figure} 
    \centering
    \includegraphics[width=0.5\textwidth,center]{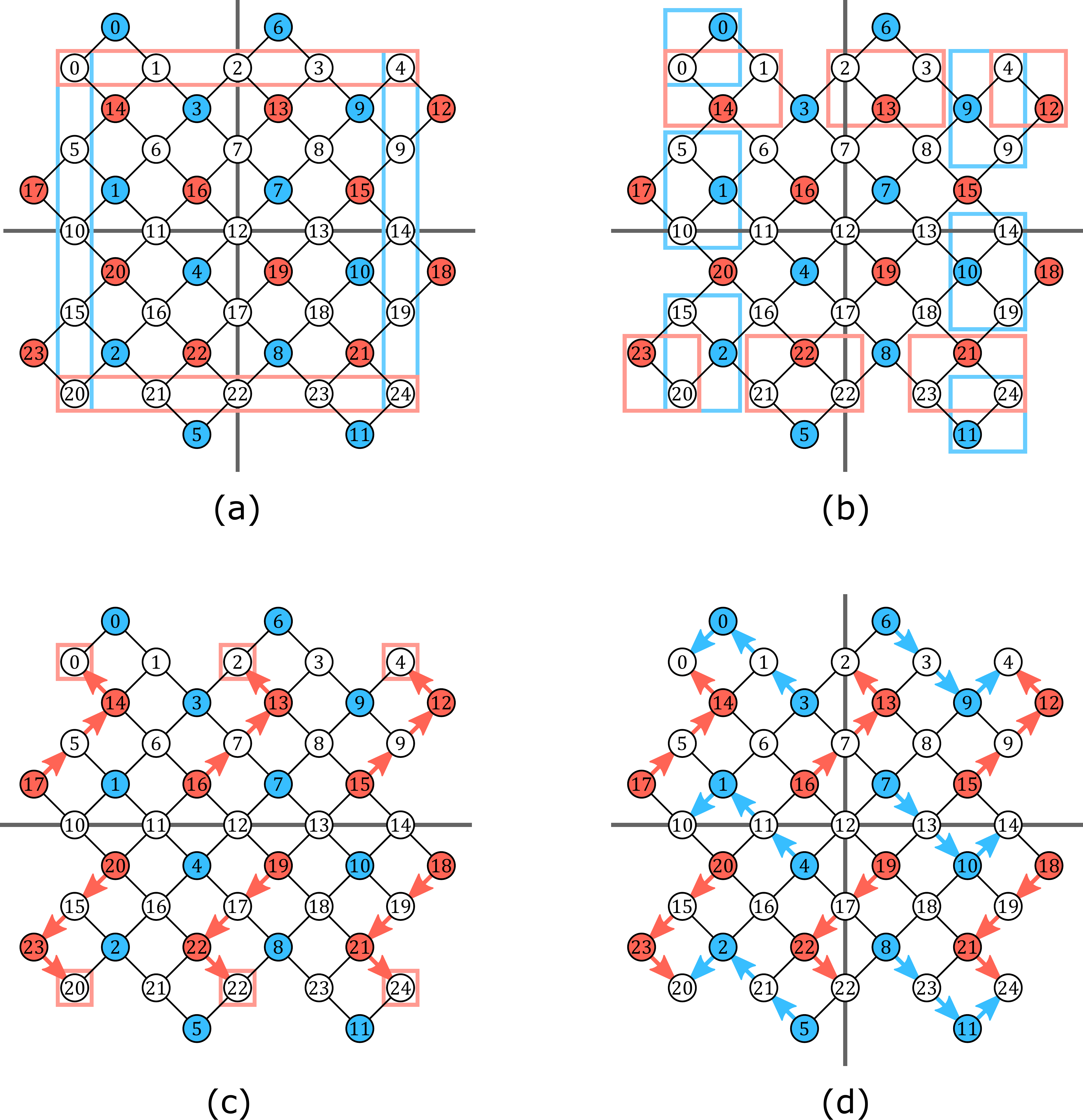}
    \caption{Figure illustrating the steps to obtain the pure error decoder used in this work. (a) The boundaries where the error chains end for the ancillas of that given colour. (b) The data qubits are grouped to the corresponding ancillas at the boundary. (c) Chains of equal length and equal distance apart for all the $Z$-ancillas, routing them to the corresponding edge. (d) Chains of equal length and distance apart for all ancillas.}
    \label{fig:pure_error_decoder}
\end{figure}

The algorithm for the PED illustrated in Fig. \ref{fig:pure_error_decoder}(d) complies with the three above-mentioned optimization targets. To understand how this PED is obtained, we recall Fig. \ref{fig:error_decomposition}, which shows that pure errors form chains of contiguous ancilla errors from the inner part to the boundaries. This means that our pure error decoder must find chains that connect all the ancilla errors to the boundaries corresponding to the appropriate logical error.
These boundaries are shown in Fig. \ref{fig:pure_error_decoder}(a). For this discussion, we will first focus on the top half ($Z$-ancillas). The full decoder can then be obtained by rotating this algorithm 3 times by 90$\deg$.

For any distance, one semi-plane has $(d^2 - 1)/4$ (6 for the example in Fig. \ref{fig:pure_error_decoder}) ancillas to be routed to the edge. As highlighted in Fig. \ref{fig:pure_error_decoder}(b), at the edge, there are $(d+1)/2$ (3) ancillas and $d$ (5) data qubits. As we need only one data qubit per ancilla, we only need $(d+1)/2$ (3) data qubits as well. For symmetry, we choose to route each ancilla to the data qubit on the boundary that are equally spaced, as shown in Fig. \ref{fig:pure_error_decoder}(c). To be invariant to translations, all $(d+1)/2$ (3) error chains are kept  equidistant when moving towards the boundary. Since we have $(d^2 - 1)/4$ (6) ancillas, each chain will be $(d-1)/2$ (2) ancillas long. By rotating this scheme 3 times by 90$\deg$, each ancilla is routed to the boundary as in Fig. \ref{fig:pure_error_decoder}(d). Combining this pattern with the numbering as shown, an algorithm to be executed in software or hardware can be derived.  This algorithm has minimized the length of the longest chains, making the hardware as fast as possible. As it turns out this means that all chains have equal lengths.

The resulting algorithm is just a series of XOR gates and can be described by the iterative formula in equation \ref{eq:pedalg} with initial step \ref{eq:pedalginit}. Here $i$  indicates the step in the algorithm, starting from the center at $i=0$ to the edge at $i = (d-1)/2-1$. $E(q_i)$ is the error on data qubit with number $q_i$, and $E(a_i)$ is the error on the  ancilla $a_i$.
\begin{align}
E(q_0) &= E(a_0)\label{eq:pedalginit}\\
 E(q_i) &= E(a_{i}) \oplus E(q_{i-1}) \label{eq:pedalg}
\end{align} 
The indices $q_i$ and $a_i$ correspond to  the data and ancilla qubits in Fig. \ref{fig:pure_error_decoder}(d) and can be calculated as:
\begin{align}
 \scriptstyle q_i = & \scriptstyle \big[\frac{d-1}{2} + r\cdot(i+1) + 1\big]\cdot\big[t\cdot d + (1-t)\big] -1 + 2\cdot c\cdot\big[d\cdot(1-t)-t\big] \label{eq:pedindexq}\\
 \scriptstyle a_i = & \scriptstyle \big[\frac{d^2-1}{4}\big]\cdot\big[1+2\cdot t\big] +\big[\frac{r-1}{2}+r\cdot i\big]\cdot\big[\frac{d+1}{2}\big] + c \label{eq:pedindexa}
\end{align}
where $t$ is either 0 or 1 for an $X$ or $Z$-chain, respectively, $r$ is the rotation of the algorithm, being $-1$ or $+1$ for left or right for $X$-chains and up or down for $Z$-chains and $c$ is the specific chain.
For example, if there is an error on $X$-ancilla 8, we take a look at $t=0$, $r=+1$ and $c=2$. If we plug these values into equations \ref{eq:pedindexq} and \ref{eq:pedindexa}, we obtain:
\begin{align}
 q_i = &\, 23+i\\
 a_i = &\, 8+3i
\end{align}
The initial step $i=0$ says that there is an error at $q_0=23$, because there is an error at ancilla $a_0=8$. Next,  there is also an error at $q_1=24$, as there is no error at ancilla $a_1=11$. This results in the same pure error as seen in Fig. \ref{fig:error_decomposition} $P$. The output of the pure error decoder is the sum of all data errors after running this iterative process over all chains on both sides of both ancilla types.

\subsection{Neural Network}
\begin{figure}[b]
    \centering
    \includegraphics[width=0.25\textwidth,center]{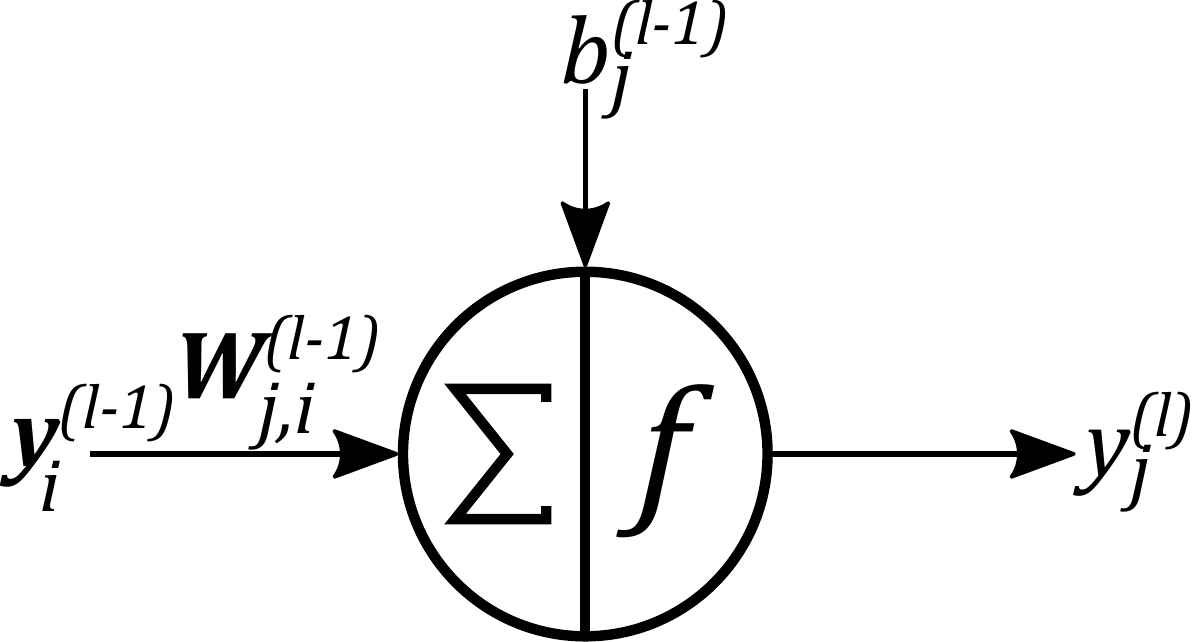}
    \caption{Illustration of a computing node in a neural network. The picture shows node $j$ in layer $l$.}
    \label{fig:nn}
\end{figure}

\begin{figure}[t]
    \centering
    \includegraphics[width=0.48\textwidth,center]{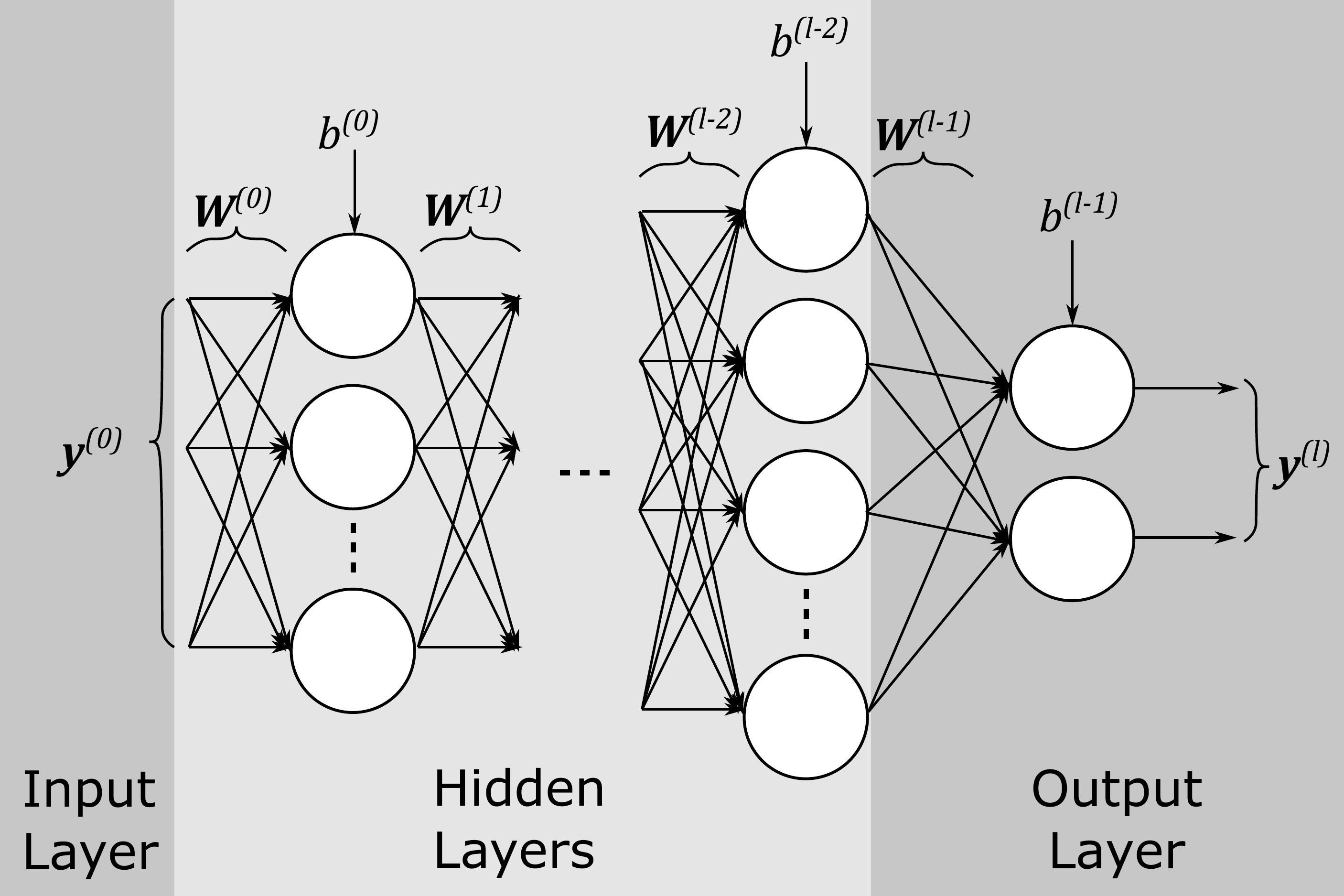}
    \caption{Illustration of a fully-connected feed-forward neural network. This work uses two hidden layers and two outputs as depicted here.}
    \label{fig:nn_full}
\end{figure}

As mentioned earlier, a neural network is used to determine the logical error made by the error estimation of the pure error decoder. The input to the neural network are the syndromes of all ancillas and the output is one of the possible logical errors. This work uses a fully-connected feed-forward neural network, which is a regular multilayered structure consisting of computing nodes. Every node in a layer is connected to all the nodes in the previous and following layer, as illustrated in Fig. \ref{fig:nn} and Fig. \ref{fig:nn_full}.


Every node sums its weighted inputs, adds a bias to the resulting sum and applies a non-linear function to the result to generate the output. This is expressed analytically as:
\begin{align}\label{eq:nodein}
    a_j^{(l)} &= \sum_i{W_{j,i}^{(l-1)}\cdot y_i^{(l-1)} + b_j^{(l-1)}}\\\label{eq:nodeout}
    y_j^{(l)} &= f\left( a_j^{(l)} \right)
\end{align}
where $y_i^l$ is the output of node $i$ on layer $l$, $W_{j,i}^{(l-1)}$ is the weight to be applied to the output of node $i$ of the previous layer ($l-1$) when contributing to the node $j$ of the  layer $l$, $b_j^{(l-1)}$ is the bias, $a_j^{(l)}$ is the accumulated output and $f(\cdot)$ is a non-linear transfer function (or activation function).
The non-linearity of the transfer function is crucial to avoid the whole neural network collapsing into a single linear layer.

A neural network always contains an output layer. As the name suggests, all the nodes in this layer produce the outputs of the neural network. The vector of inputs is sometimes called the input layer. However, as can be seen in Fig. \ref{fig:nn_full}, this layer does not contain any nodes. If more layers are used between the input layer and the output layer, they can not be directly observed, and are hence called hidden layers.

Even though a single hidden layer is enough to map any function \cite{Hornik1989}, having multiple layers reduces the number of nodes needed in each layer. Previous work showed us that two hidden layers performs better than a single layer in terms of decoding accuracy \cite{Varsamopoulos2018_2}. Since adding another layer did not yield any significant improvement, we will focus on two hidden layers in this work.

The number of inputs of every node depends on the number of nodes of the previous layer, except for the first hidden layer. In this layer, the number of inputs is equal to the number of ancilla qubits, i.e.~$d^2-1$.

The number of nodes in the output layer depends on the classification scheme. One can use the classification scheme shown in Fig. \ref{fig:outputsped}(b), using four nodes to represent the different errors. This can either be no error, called a logical identity $I$, or one of the logical $X$, $Y$ or $Z$ errors. These logical errors represent the logical difference between the pure error decoder output and the actual data errors. However, this can again lead to different output nodes competing and deciding independently. For this reason, we choose only two output nodes, one for signaling a logical $X$ error and the second one for a logical $Z$ error. This has the added benefit of reducing the number of output nodes and thus the weights and size of the neural network, whilst still keeping the four output classes as no or both $X$ and $Z$ give $I$ and $Y$ respectively.

Although a more complex architecture, such as recurrent and convolutional neural networks, would be very suited for this application, we opted for the simplest implementation of a fully connected feed-forward neural network as a first step towards the hardware implementation of NN QEC decoders in hardware. The obtained results will be the basis for  future extensions to more complex cases.

\section{Methods for simulation and training }\label{sec:Setup} 

\begin{figure} 
        \centering
        \includegraphics[width=0.45\textwidth,center]{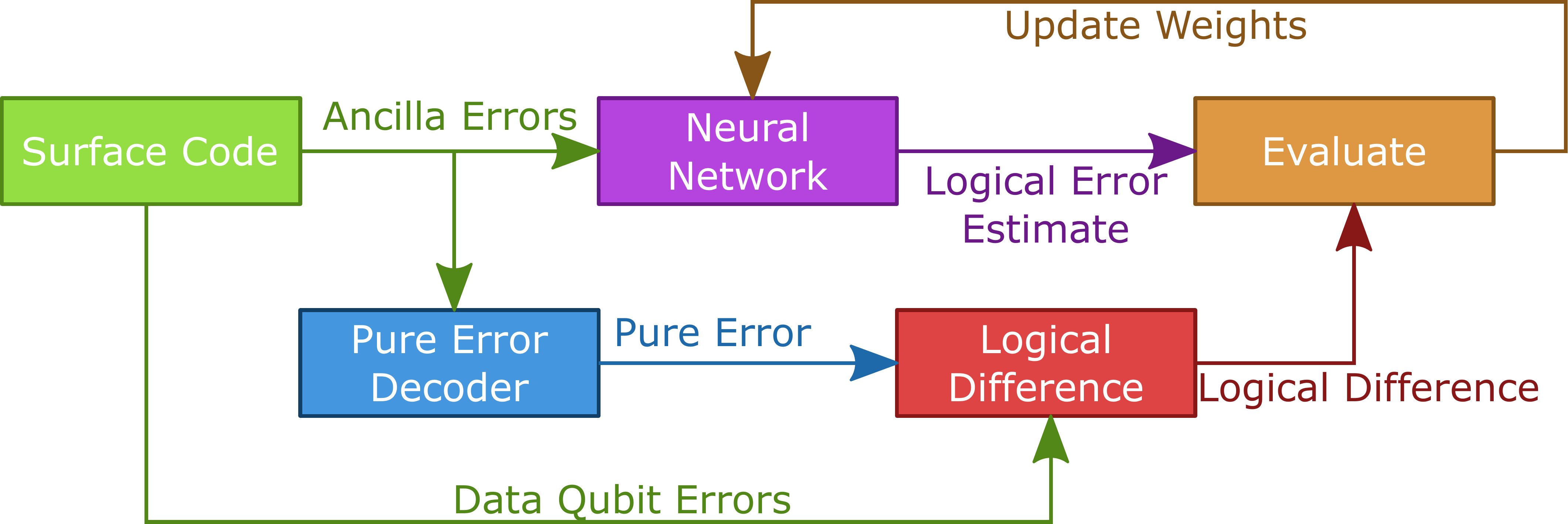}
        \caption{Overview of the simulation setup used in this work.}
        \label{fig:setup}
    \end{figure}

	Before delving into the design and optimization of the different parameters of the neural network, the details about the simulation infrastructure are first described. The flow of the simulation setup is shown in Fig. \ref{fig:setup}. First, to generate a realistic error pattern for both NN training and evaluation, the data qubit errors are sampled using the depolarizing error model as discussed in the next section. Those are then fed to the surface code simulator to obtain the corresponding error syndrome. The error syndrome is passed to the pure error decoder, which returns the pure error. The pure error is compared with the actual data qubit errors and the logical difference between the two is saved as the target output for the neural network. The error syndrome is also given to the neural network, which produces a logical error estimate. By comparing such an estimate to the target logical difference, the correctness of the estimation can be derived, which can then be used to assess during the training of the NN or to evaluate its performance.
	
	\subsection{Sampling}
	Having no recurrency in our decoder means that the depolarizing error model without measurement errors is chosen for this work, because our decoder would not be able to decode the time-dependent errors deriving from the measurement errors. The error model is implemented by applying a random physical error on each data qubit in every cycle chosen among a $X$, $Y$ or $Z$ error with equal probabilities $p/3$. 
	This sampling is done on the fly just before the neural network is run, without pre-generating a dedicated training and testing data set.
	
	Prior work \cite{Varsamopoulos2018_2} investigated the optimal way of generating such a data set without overfitting. They sampled a large number of data qubit error configurations and recorded the resulting syndrome and logical error of the pure error decoder. The neural network was then trained on this syndrome data set with the target output being the corresponding logical error distribution, until it reached a certain accuracy on the training data set. However, after generating such a data set for distances larger than 5, the number of possible syndromes becomes so large that only a single data point is sampled per syndrome. Since the neural network is then not trained on a representative error distribution, there is no guarantee of generalization.
	
	Instead, thanks to our simulation setup, this work keeps sampling new data on the fly without a limit in the size of the data set. The training procedure itself will then average out all of these points and the neural network will learn the error distribution. Overfitting is avoided as all the data will be new, and the neural network will only benefit from training longer. In addition, since the data set is uncorrelated, the resulting logical error rate at the end of the training will always represent the performance of the neural network.
	
	The work in \cite{Varsamopoulos2018_2} found that training at a certain physical error rate will optimize the performance of the neural network at that physical error rate. Because we want to optimize the performance at the $p_{th}$, we sample at the physical error rate corresponding to the $p_{th}$ of the MWPM algorithm for that distance. 
			
\subsection{Training and testing}
    The training is done using the ADAM optimizer \cite{Kingma2014} with a batch size of 4992. As we have no finite data set to optimize for, we trained the neural network for 300,000 batches. This results in a total data set of $\approx 1.5\cdot10^9$ for each training. Fig. \ref{fig:training_d9_tot} plots the logical error rate during training per iteration of 2000 batches on the largest used neural network, showing the performance saturation after 150 iterations.
    
    The testing after training is done similarly to training. We again run 2000 batches to obtain the desired statistical accuracy. This is done for several logarithmically spaced values in a range between 0.03 and 0.3. As can be seen in Fig. \ref{fig:blossomslope}, this range includes the pseudo-threshold, the decoder threshold, and clearly shows the slope difference. To obtain the slope, a fit is performed using the model in equation \ref{eq:model}, and to obtain the $p_{th}$, we interpolate the two values above and below the $ler=per$ line in the logarithmic domain. The reported variance used in the confidence interval is the sum of the variances of these two points.
    For the simulations that include quantization, this process is repeated for every combination of quantization levels and regularization levels.
    
    All training and testing was done on custom written code in cpp and Cuda which was ran on NVIDIA Tesla K40 GPUs over a span of a couple of months. All code is available at \cite{DOI}.

\subsection{Cost function for quantization}
\begin{figure}
        \centering
        \includegraphics[width=0.4\textwidth,center]{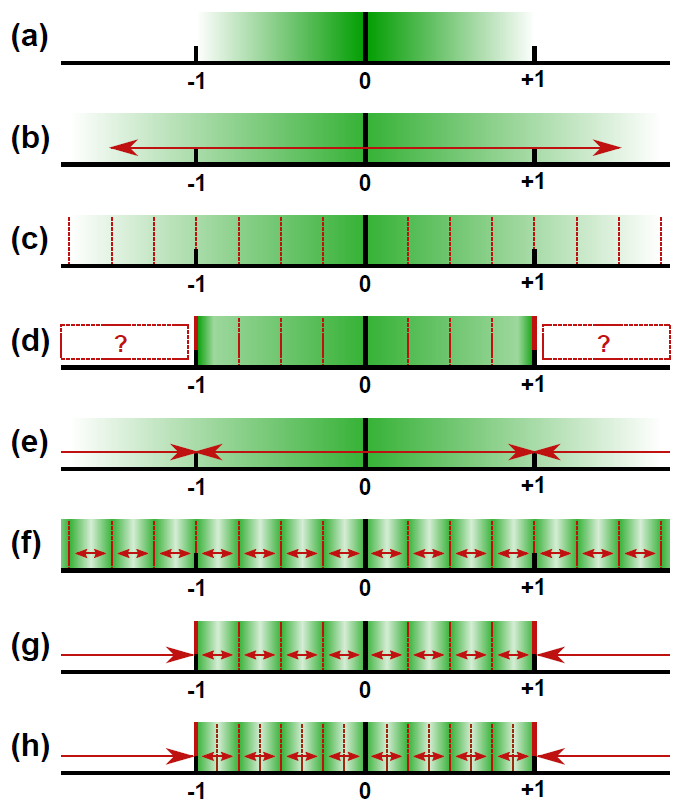}
        \caption{Method adopted for training to optimize the representation of the weight using fixed point rather than floating point. This involves using weight regularization to push the weights towards zero and towards the nearest quantization level. The weights and outputs are then quantized after training.}
        \label{fig:quantization}
    \end{figure}

\begin{figure} 
    \centering
    \includegraphics[clip,trim=0.5cm 0.5cm 1cm 0cm,width=0.50\textwidth,center]{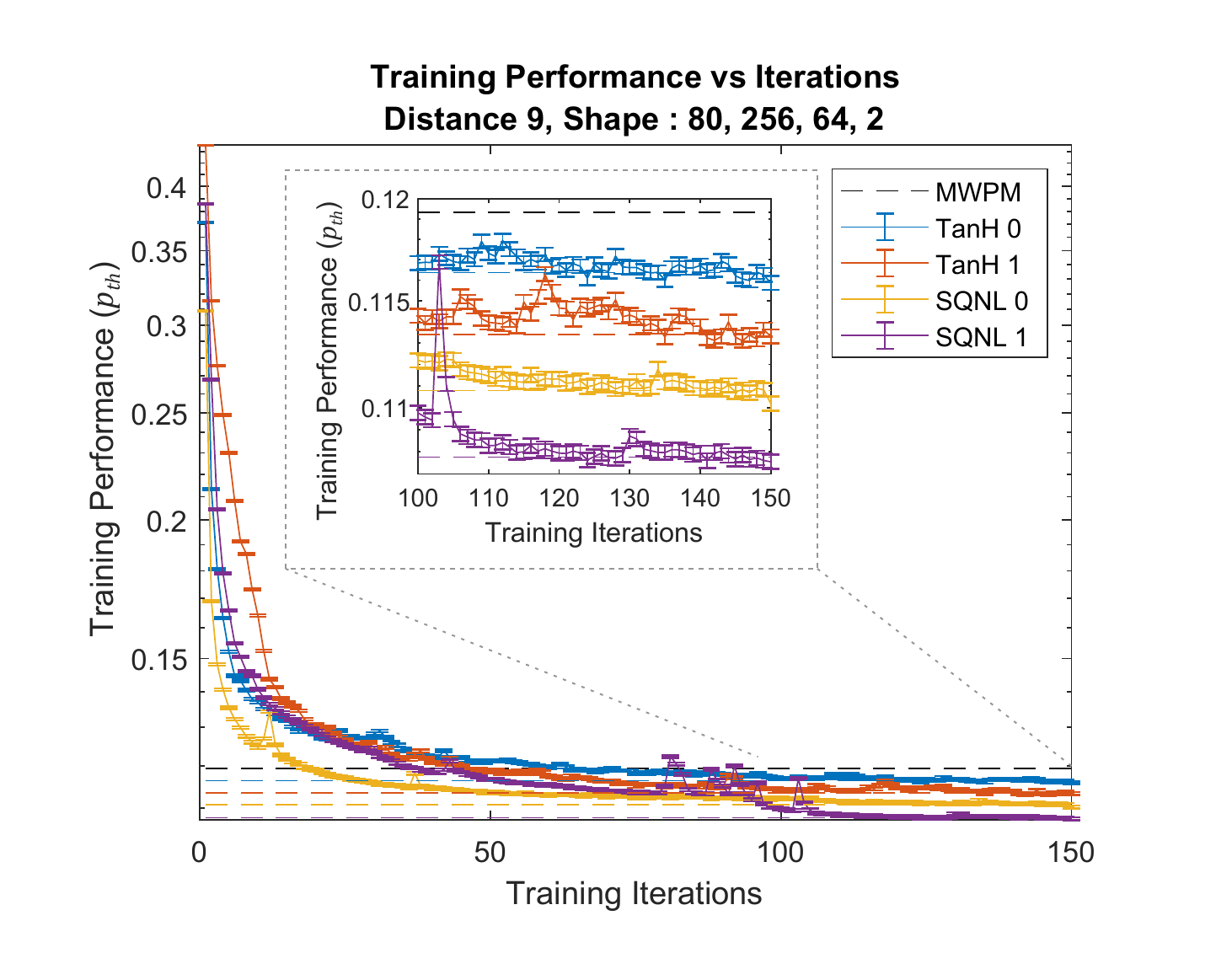}
    \caption{Logical error rate during training for a distance 9 code with the maximum neural network used in this work (hidden layer sizes: 256 and 64), comparing to the MWPM decoder the final error rate for different transferfunctions (TanH and SQNL) and for the use of the rotational symmetry (0 or 1). The error bars show a confidence interval of 99.9\%.}
    \label{fig:training_d9_tot}
\end{figure}

    Due to the targeted hardware implementation, some additional regularization terms are added to the typical mean-squared-error cost function used during the NN training. The  process is illustrated in Fig. \ref{fig:quantization}. Usually, the weights are randomly initialized in a certain range [\ref{fig:quantization}(a)]. During training, the weights expand outwards [\ref{fig:quantization}(b)] \cite{Bishop1995}, which usually is not a  problem for  weights using the floating-point representation. However, if we want to quantize those weights to a set of discrete levels [\ref{fig:quantization}(c)], issues arise when limiting  the number of bits used in quantization. If, for example, all the weights are quantized  between -1 and +1, all weights outside this region are clipped to -1 and +1 [\ref{fig:quantization}(d)]. To push the weights towards 0 during training, we can add the sum of all squared weights $|w|^2$ to the cost function [\ref{fig:quantization}(e)] \cite{Bishop1995}. This will push less important weights to zero and decrease the average size of all weights. Another problem is that before quantization all the weights are uniformly distributed between the $\pm1$ range. To minimize the quantization error, we can also try to push the weights towards certain quantization levels during training [\ref{fig:quantization}(f)]. This can be done by adding the sum of the squares of the difference between every weight and the nearest quantization level $|w - w_q|^2$. Combining the two [\ref{fig:quantization}(g)] results in the following cost function: 
    
    \begin{equation}\label{eq:cost}
        \sum\left(y-t\right)^2+r\cdot\left(\sum|w|^2+\sum|w-w_q|^2\right)
    \end{equation}
    where $y$ is the output, $t$ is the target output, $w$ is the value of the weight and $w_q$ the quantized weight.
   The additional scaling term $r$ decreases the influence of the weight regularization compared to the output error. 
    
    To further decrease the quantization error, a different number of quantization bits can be used to sample the weights than is used for the regularization. An example where we sample with an additional bit is shown in Fig. \ref{fig:quantization}(h). The final reported performance is the optimum over all possible regularization bits. Finally, because we use two's complement signed fixed-point numbers, the discussion above should be in the range $[-1, 1 - 1/2^{b-1}]$, where $b$ is the number of bits.\\

\section{Decoding Performance Results}\label{sec:Decoding_Results} 

The main objective of this work is to minimize the complexity of the neural network used in the HLD, while still obtaining a competitive decoding performance. Reducing the complexity implies a reduction in the amount of free parameters. For instance, this can be done by reducing the size of the neural network, or by constraining the architecture and weights. Constraining should be done with care \cite{Sutton2019}, but can reduce the size while still improving the performance \cite{LeCun1989}. An example are CNNs, where the connectivity is limited and weights are reused.

In this work, we focus on four tuning knobs that influence both the decoding performance and the hardware cost: the rotational symmetry, the transfer functions, the layer sizes and the number of bits used for quantization. When looking at these parameters, we will compare the obtained decoder slope and the pseudo-threshold. First, the influence of rotational symmetry is determined. Next, different transfer functions are compared. These results are then used in a layer-size sweep and in a bit-width sweep for the quantization.

\subsection{Rotational Symmetry}
This work focuses on fully connected neural networks, thus leaving  exploration of CNNs for future work. However, the rotational symmetry of the surface code and our PED can be investigated. The weights of the neural network can be copied and rotated four times. If we then use an initial neural network with a quarter of the size, the number of independent weights is divided by four, while still keeping the same total amount of weights and  connectivity. This will both reduce the hardware cost by reducing on-chip memory, and ease the optimization during training.

In order to compare the performance difference between rotating the neural network or not, we ran simulations for all different distances, transfer functions (TanH, ReLU, SQNL, see next section),  and layer sizes (from 4 to 256). For all these simulations, both a version with and without rotational symmetry is trained. The $p_{th}$ and slope have been extracted and the averages over the different transfer functions and layer sizes are presented in the top two rows of Table \ref{tab:rot_tf}. The average is calculated by taking the geometric mean of the ratio between including and excluding symmetry for every configuration. If the ratio is larger than 1, an improvement is found by rotating. Since the performance for some configurations  was too low, resulting in an undefined slope or $p_{th}$,  those configurations were excluded from the computed average.

We find that including rotational symmetry has a positive effect on both the $p_{th}$ and the slope for all distances. This is mainly attributed to the training having an easier time finding an optimum and is in line with the results in \cite{Wagner2019} for a Toric code. The improvement  increases for larger distances, which is likely due to the larger neural networks performing better and  benefiting more from the weight regularization. This is in line with the results found on layer sizes later in this section. 

\subsection{Transfer functions}
Different transferfunctions will have different hardware costs and decoding performance, with this work will focus on, in descending hardware cost, hyperbolic tangent (TanH), SQuared Non-Linearity (SQNL) and Rectified Linear Unit (ReLU), defined as:
\begin{align}
\label{eq:tanh}
\mathrm{TanH}(x) &= \;\;\;\frac{e^{x}-e^{-x}}{e^{x}+e^{-x}} \\
 \label{eq:relu}
 \mathrm{ReLU}(x) &= 
 \begin{cases}
 \;0, & \mathrm{for} \; x < 0\\
 \;x, & \mathrm{for} \; x \geq 0\\
 \end{cases}\\
 \label{eq:sqnl}
 \mathrm{SQNL}(x) &= 
 \begin{cases}
 \;-1, & \mathrm{for} \; x < -1\\
 \;2x + x^2, & \mathrm{for} \; \; -1 \leq x < 0\\
 \;2x - x^2, & \mathrm{for} \; 0 \leq x \leq 1\\
 \;1, & \mathrm{for} \; x > 1\\
 \end{cases}
\end{align}
For simplicity, every node will use the same transfer function.

The lower half of Table \ref{tab:rot_tf} compares the performance for different transfer functions similarly to the comparison for  the  rotational symmetry. The more computationally expensive hyperbolic tangent is taken as a reference.
The average performance difference of the ReLU over the TanH is not significant. Since the ReLU is much cheaper in hardware implementation (not requiring any exponential),  it is attractive even for roughly equal performance.

The SQNL does show  a large improvement, up to 31\%. This, in combination with the simplicity of the required hardware, make SQNL  the preferred choice.

The transfer functions show the same performance increase for larger distances, similar to the rotational symmetry results. This strengthens our belief that this trend is due to the need for larger neural networks for larger distances.

Based on these results we limit the search space  for the follow-up analyses by adopting  rotated neural networks with the SQNL function. The results  with the other options are available in the supplementary materials \cite{DOI}. We also limit ourselves to the $p_{th}$ performance. 

\begin{table}
 \centering
 \caption{Average pseudo-threshold and slope improvements when comparing the use of rotational symmetry and the different transfer functions. The ratio of the geometric means of the performance over different layer sizes (excluding cases when the parameter is undefined)  is reported. }
\begin{tabular}{c c c c c c c}
    \hline
        \hline
    \multicolumn{2}{c}{Performance} & \multicolumn{5}{c}{Distance} \\
    \multicolumn{2}{c}{Comparison} & 3      & 5      & 7      & 9 & All\\
    \hline \rule{0pt}{1\normalbaselineskip}
    \multirow{2}{*}{$\displaystyle\frac{\mathrm{Rotated}}{\mathrm{Unrotated}}$}$^\text{a}$
    & $p_{th}$  & 1.0133 & 1.0320 & 1.0563 & 1.0850  & 1.0411  \\
    & Slope     & 1.0042 & 1.0107 & 1.0290 & 1.0546  & 1.0207  \\
    \\
    \multirow{2}{*}{$\displaystyle\frac{\mathrm{SQNL}}{\mathrm{TanH}}$}$^\text{b}$
     & $p_{th}$ & 1.0291 & 1.1637 & 1.2830 & 1.3154 & 1.1703 \\
     & Slope    & 1.0021 & 1.0552 & 1.1378 & 1.2137 & 1.0822 \\
    \\
    \multirow{2}{*}{$\displaystyle\frac{\mathrm{ReLU}}{\mathrm{TanH}}$}$^\text{b}$
     & $p_{th}$ & 0.9771 & 0.9957 & 1.0177 & 1.0387 & 1.0008 \\
     & Slope    & 0.9936 & 0.9978 & 1.0127 & 1.0605 & 1.0101 \\
     \hline
        \hline
        \multicolumn{6}{l}{$^\text{a}$Averaged over all transfer functions}\\
        \multicolumn{6}{l}{$^\text{b}$Averaged over both rotated and unrotated configurations}
    \end{tabular}
 \label{tab:rot_tf}
\end{table}

\subsection{Layer Sizes}
As discussed in section \ref{sec:Decoder}, this work  uses two hidden layers and an output layer with two nodes, as there are two outputs, thus using the  neural network shown in Fig. \ref{fig:nn_full} for $l=3$. We assume that reducing the number of nodes in any of the hidden layers will reduce the hardware cost but, as shown in the following, will degrade the decoding performance due to the reduced  computational power of the neural network.

\begin{figure}
    \centering
    \includegraphics[clip,trim=1cm 0.5cm 1cm 1cm,width=.5\textwidth,center]{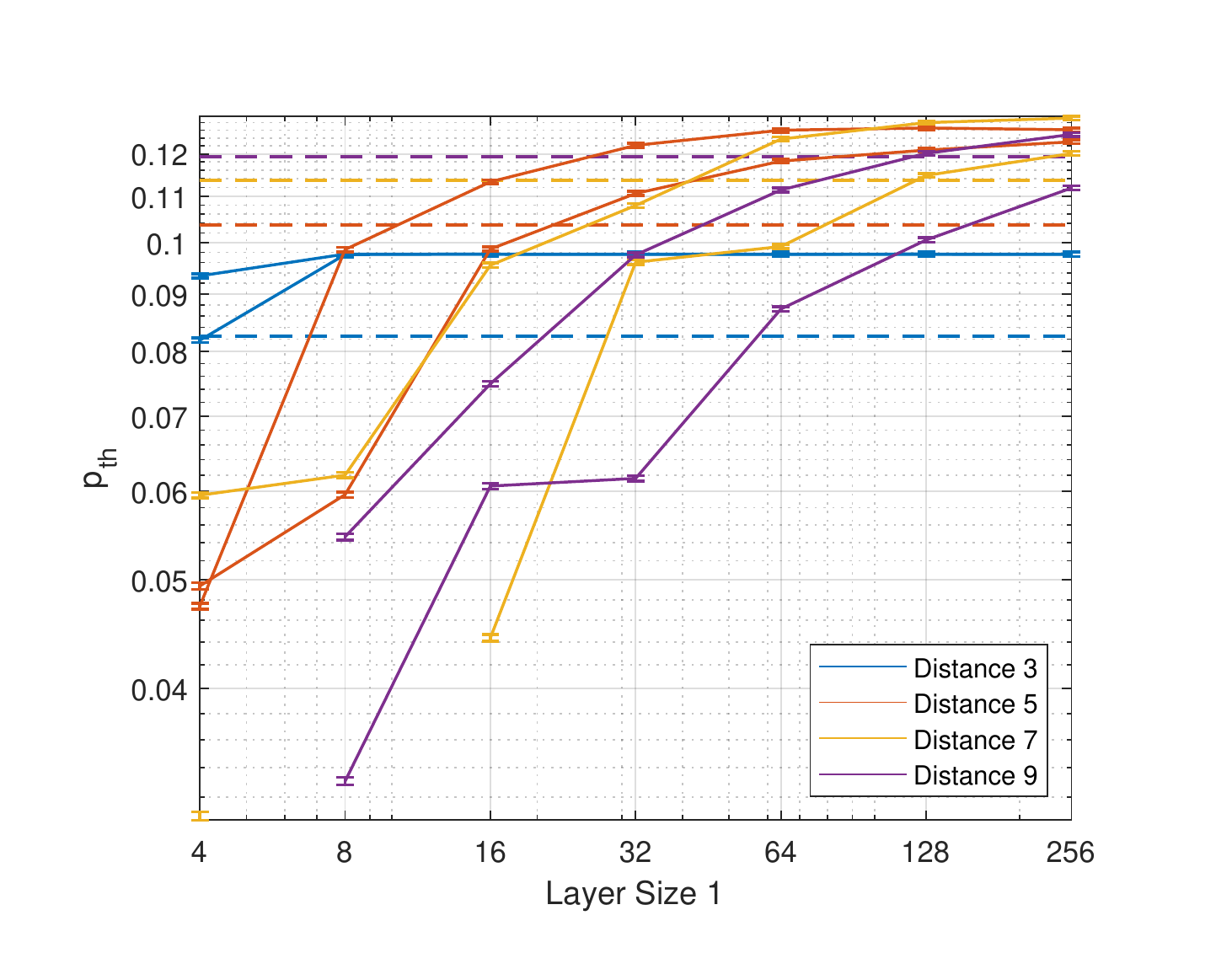}
    \caption{The $p_{th}$ performance for the smallest four surface code distances as a function of  the number of nodes used in the first hidden layer. For each distance, we report the results for a second hidden layer with 64 nodes (line with higher $p_{th}$)  and with 4 nodes (line with lower $p_{th}$). The dashed lines show the MWPM $p_{th}$. The error bars represent a confidence interval of 99.9\%.}
    \label{fig:all_layersizes}
\end{figure}

A summary of the layer size sweep  is shown in Fig. \ref{fig:all_layersizes} by plotting $p_{th}$ as a function of the number of nodes in the first hidden layer. The achieved slope is not shown in this subsection and the next, but the correlation between $p_{th}$ and the slope is shown at the end of this section in Fig. \ref{fig:slope_vs_pth}. For each distance, we report the results for a second hidden layer with 64 nodes (line with higher $p_{th}$)  and with 4 nodes (line with lower $p_{th}$). All other investigated second layer sizes lie in between these two cases. 

All lines in Fig. \ref{fig:all_layersizes} show  a similar trend, saturating   to a maximum performance with increasing  layer sizes. This maximum $p_{th}$ increases for larger distances, and, as expected before, a larger distance requires a larger neural network for the same $p_{th}$. The largest tested neural network decoder (256 and 64 nodes in the first and second hidden layer, respectively)  is enough to outperform MWPM, whose performance is indicated by the dashed lines. For the smallest distance of 3, almost any neural network reaches maximum performance. For distance 9, the maximum is not yet visible, indicating   that future research should include larger neural networks.

The  effect of the second layer size  is shown as the difference between the higher and lower line for each distance. More detailed data are included in \cite{DOI}. The first hidden layer size has a stronger impact on performance than the second  layer, although this is stronger  for larger distances.

\subsection{Quantization}
Using a fixed-point representation for the data in the NN instead of a floating-point representation can significantly save  hardware cost, but at the price of decoding performance. Since there is no strategy to determine the optimal number of bits in the fixed-point representation \cite{Verhelst2020}, we explore how performance varies for different quantization levels. Ideally,  the optimum number of bits for each node depends on the number of input nodes and the domain of the transfer function, and in principle can be different for node output, weights and biases. However, for simplicity of the study, we adopt the same  number of bits to represent the outputs, the weights and the biases in all layers, except for the 1-bit global inputs and outputs of the NN. 

\begin{figure}
    \centering
    \includegraphics[clip,trim=1cm 0.5cm 1cm 1cm,width=0.5\textwidth,center]{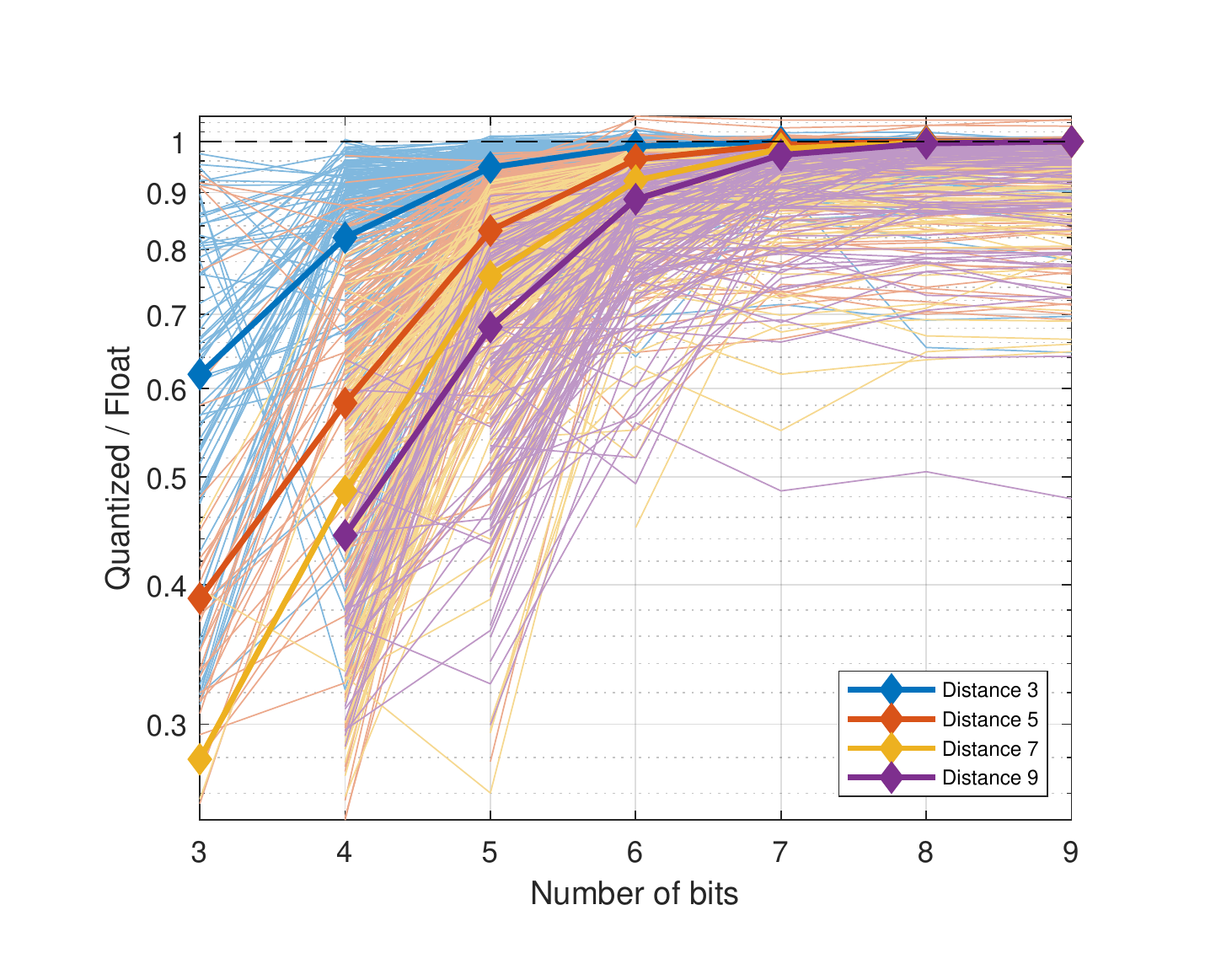}
    \caption{The $p_{th}$ performance degradation of the quantized neural networks with respect to their floating-point counterparts. All configurations are shown in lighter colours. The thicker lines represent the average performance degradation.}
    \label{fig:q_all_norm}
\end{figure}

To see the effect of quantization, we compare the performance of the neural network before and after quantization. This is done by dividing the quantized (fixed-point) $p_{th}$ by the floating-point $p_{th}$. The floating point performance before quantization does include the regularization term that attracts the weights to certain quantization levels. These results are shown in Fig. \ref{fig:q_all_norm}, plotting the performance degradation due to  the quantization for all different configurations of layer sizes and quantization levels used in weight regularization. For layer sizes this is between 4 and 256 in the first hidden layer and between 4 and 64 in the second hidden layer. The weight regularization levels are varied between 4 and 256 levels (2 and 8 bits). The $x$-axis shows the amount of bits used for representing the data during evaluation. 

A couple of trends can be observed. First, 9 bits are enough  for most configurations to reach floating point performance (a value of 1). Next,  larger distances  need more bits for the same performance degradation. These trends are captured in the thick lines as a cumulative average performance degradation. These lines represent the average degradation in performance when going down to less bits. First all point at 9 bits are averaged for a certain distance. Then every following step down in bits, we subtract the average degradation of all lines in that distance. 

From the data \cite{DOI} we also see that we need more bits to reach the MWPM performance for larger distances. For increasing distances we need 3, 4, 5 and 7 bits. One explanation could be due to the relation between layer sizes and quantization discussed at the start of this section. This would indicate that more research into   quantization dependent on layer size is needed. However, another option would be to look into CNNs, as that also decreases the number of inputs per node.

\subsection{Comparing decoding performances}
Table \ref{tab:results} combines all the results discussed so far. It compares the maximum $p_{th}$ and slope found in this work to the neural network decoders in \cite{Varsamopoulos2018_2} and the MWPM decoder. The data in the table is limited to the use of the rotational symmetry and the SQNL activation function, and to a layer size of 256 and 64 for the first and second hidden layer, respectively.
Even with neural networks  smaller than  in \cite{Varsamopoulos2018_2}, we obtain slightly higher pseudo-thresholds, thus confirming the validity of  our training method. It also means that our choice for a simpler transfer function does not affect the performance. Finally, we see that quantizing (at least to 8 or more bits) does not result in any significant degradation in performance.\\

Interestingly, a distance-3 decoder requires only a very small neural network and a few bits. Because, as shown before, the added symmetry improves performance, an interesting possibility is using CNNs \cite{Ni2018} or distributed decoders \cite{Varsamopoulos2019} based on distance-3 kernels, thus requiring simpler hardware, allowing the decoding of larger distances and even increasing the decoding performance.

\begin{table}
 \centering
 \caption{Decoding performance of this work compared to the MWPM decoder and the work of \cite{Varsamopoulos2018_2}. Results are limited to the use of the rotational symmetry and the SQNL transfer function, for both the best floating-point performance and the best performing fixed point neural network. We also report the minimum number of bits needed to obtain a $p_{th}$ higher than that of the MWPM decoder.}
 \begin{tabular}{c c c c c c}
 \hline
        \hline
  Performance & Used & \multicolumn{4}{c}{Distance} \\
  Parameter & Decoder & $3$ & $5$ & $7$ & $9$ \\
  \hline \rule{0pt}{1\normalbaselineskip}
  &MWPM & 0.08251 & 0.10372 & 0.11368 & 0.11932 \\
  &\cite{Varsamopoulos2018_2} & 0.09815 & 0.12191 & 0.12721 & 0.12447\\
  &Float & 0.09769 & 0.12657  & 0.12917  & 0.12490  \\
  \multirow{-4}{*}{$p_{th}$}
  &Fixed & 0.09781  & 0.12637  & 0.12934  & 0.12430  \\
  \\
  &MWPM & 1.856 & 2.723 & 3.601 & 4.496 \\
  &Float & 1.886 & 2.869 & 3.812 & 4.663 \\
  \multirow{-3}{*}{Slope}
  &Fixed & 1.894 & 2.866 & 3.820 & 4.667 \\
  \\
  \multirow{-1}{*}{Min. Bits}
  & Fixed & 3 & 4 & 5 & 7 \\
  \hline
        \hline
 \end{tabular}
 \label{tab:results}
\end{table}
 
\begin{figure}
    \centering
    \includegraphics[clip,trim=1cm 0.5cm 1cm 0.5cm,width=0.5\textwidth,center]{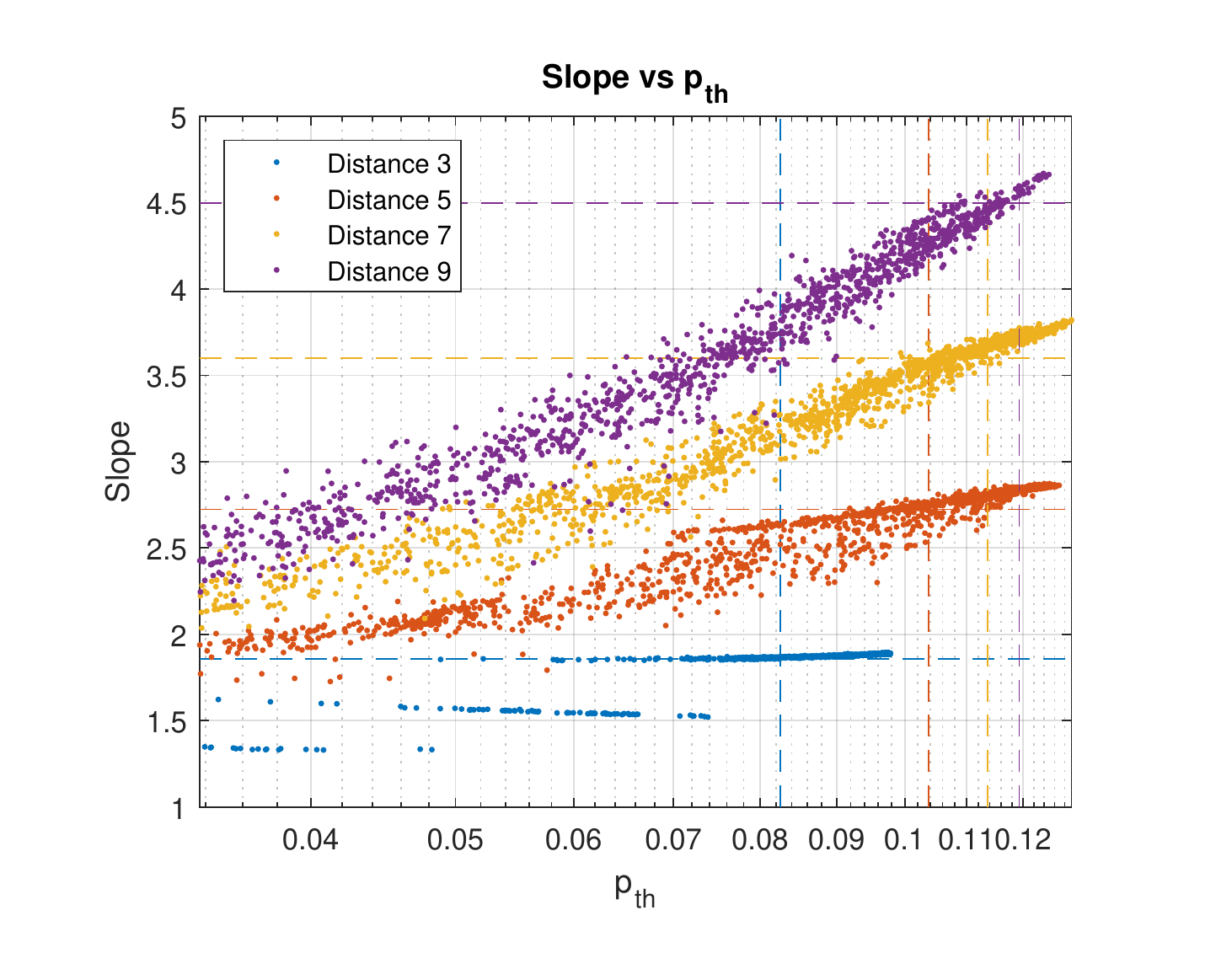}
    \caption{Correlation between the slope and pseudo-threshold for the four distances. Every dot represents a different neural network configuration (rotated with SQNL function and all layer sizes). The dashed lines indicate the slope and $p_{th}$ of the MWPM decoder.}
    \label{fig:slope_vs_pth}
\end{figure}

 Finally, an overview of all decoders is shown in Fig. \ref{fig:slope_vs_pth}. In this figure every decoder is depicted with a dot in the color of the corresponding distance, and the dashed lines represent the performance of the MWPM decoder. The correlation between the slope and the $p_{th}$ is quite apparent. The main takeaway from this plot is that a decoder on a larger distance can have a worse slope   than a decoder on a smaller distance. Thus,  if the decoder can not achieve sufficient performance, it might not be economical to go for a larger distance. For example, a decoder could be too large to be co-integrated with the qubits, or consume too much power, or have too large delay  to keep up with the generated data. For this reason, the next section will analyze the estimated hardware cost.
 
\section{Hardware costs}\label{sec:Hardware_Results}

	\subsection{Hardware Estimate}\label{sec:setuphardware}
	As we want to find out the minimal achievable delay  and we do not need any memory for recurrency inside of the neural network, we chose a fully parallel combinatorial implementation. Some flip-flops would be needed to store  the input and  outputs of the neural network during inference, and the weights and the biases would be stored in some external RAM. However, for illustrating  the effects of solely the neural-network logic on the hardware, the cost in area and power of those memories and the memory access are not included in the following estimates.
	
	Although one could expect that the absence of a clock and flip-flops  would decrease the power consumption, this is  not always the case \cite{Parhami2012}. A fully combinatorial circuit will propagate any glitch to the outputs, thereby increasing the amount of charging and discharging of the  capacitance of the digital cells and the  interconnect parasitics. Instead,  these glitches would not be able to propagate in a more pipelined approach. In combination with  clock gating , this could even decrease the power consumption \cite{Parhami2012}, although   pipelining would increase the area and delay.
	
\begin{figure}
        \centering
        \includegraphics[width=0.50\textwidth,center]{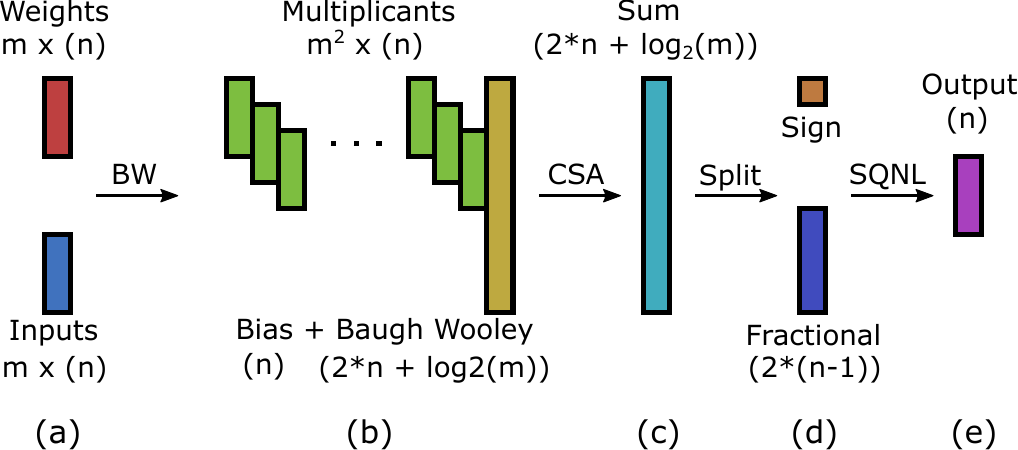}
        \caption{Schematic of the hardware in one node. First $m$ $n$-bit inputs are combined with their corresponding weights (a) into the multiplicants needed for the Baugh-Wooley (BW) multiplication scheme (b). These multiplicants are then added in a Carry Save Adder (CSA) tree (c). Finally the result is split (d) and passed through the SQNL block to obtain the $n$-bit output (e).}
        \label{fig:nodehardware}
    \end{figure}

	The data flow  in each node is depicted in Fig. \ref{fig:nodehardware}. First, all the multiplicants are calculated using the Modified Baugh-Wooley 2's complement method \cite{BaughWooley73, Parhami2012} (b). These are then summed in a Wallace carry-save adder tree \cite{Wallace1964} (c). Finally, to implement the SQNL function (equation \ref{eq:sqnl}), the resulting fractional part of the sum (d) is passed through a squaring unit \cite{Parhami2012} and added to or subtracted from a bit-shifted sum depending on the sign (e). This process is different for input nodes and output nodes. Each input is only one unsigned bit, and thus the Baugh-Wooley method is replaced by a simple AND operation. The outputs are also just one bit, thus only the sign of the sum is needed.
	
	A hardware description of the resulting digital circuit is fed to the GENUS synthesizer using the CMOS standard static library form the adopted TSMC 40-nm CMOS process to obtain the circuit schematic. To efficiently obtain the hardware estimates for all the  neural network configurations, only the individual nodes are synthesized. The resulting delay, area and power from the individual nodes are then summed to obtain an estimate for the total neural network. 
	
	To estimate the delay of the NN, we extract from the synthesis the critical path of the node, which is equal to the critical path of the respective layer as every node in a layer is the same. The total delay  is simply obtained by  summing the critical path delays of the layers. This estimate does not include the additional delay due to the interconnect between layers.
	
	The area of each node is estimated as the sum of the area of the digital cells in the node, scaled by a fixed fill factor of $1$. Since all the nodes are equal, the total area is just the sum of the node areas, as we assume that the top-level layout can be quite efficient and does not consume any additional area. 
	
	The power is defined as the average energy needed per decoding cycle of 440 ns. To get a more accurate power estimate, we perform transient simulations in Cadence Virtuoso of the nodes. This will include the extra power consumption due to the propagation of glitches. The energy is then averaged over 100 cycles. The simulations have an input activity factor of 50\%.
	
	\subsection{Results}\label{subsec:HardwareResults}
\begin{figure*}
    \centering
    \includegraphics[clip,trim=2.2cm 0.5cm 2.2cm 1cm, width=\textwidth,center]{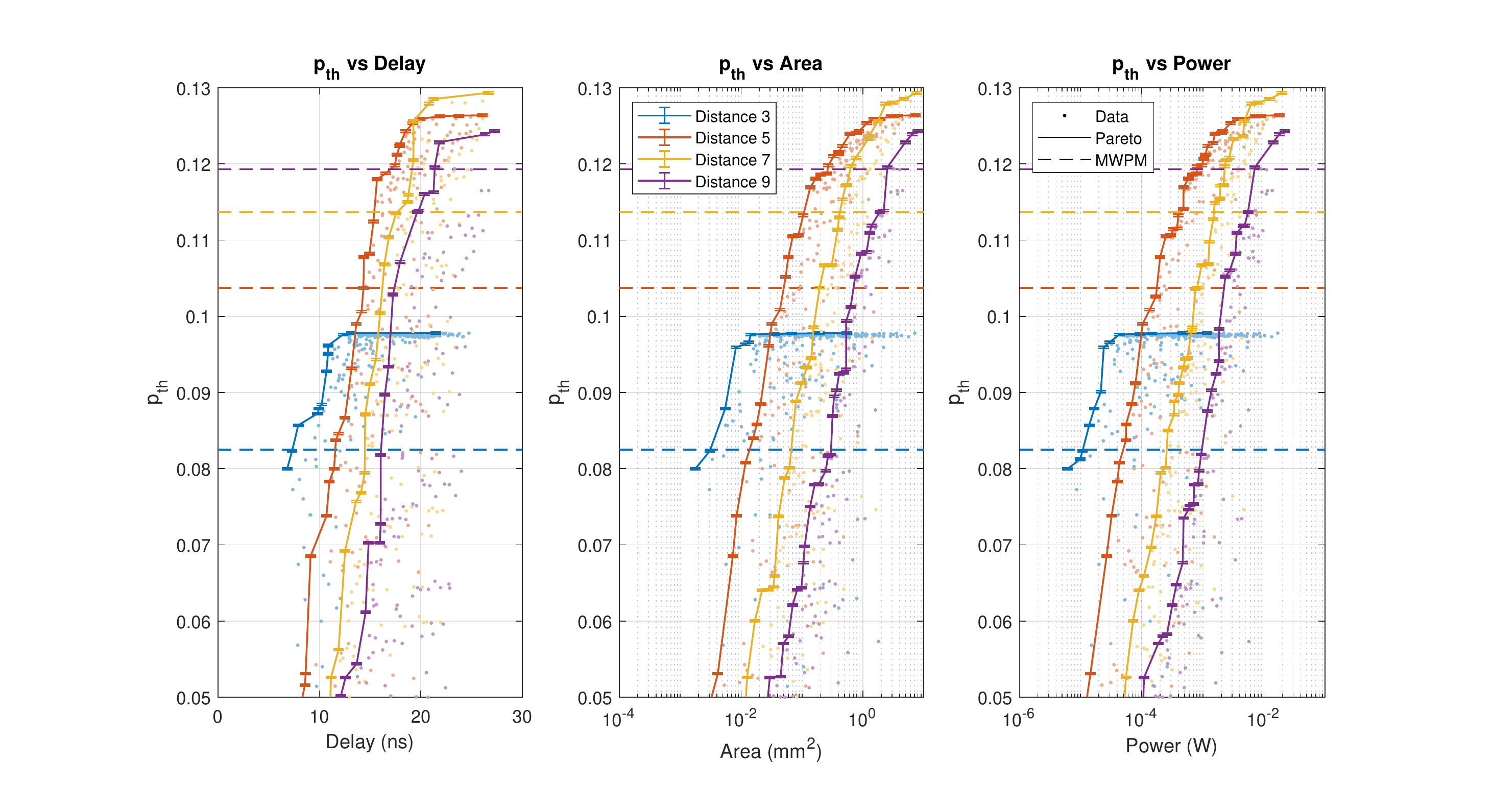}
    \caption{Pseudo-threshold of the quantized neural network decoders versus their estimated hardware cost. The figure shows the Pareto front of the different distances and compares them to the MWPM decoder. The error bars represent a confidence interval of 99.9\%}
    \label{fig:Pareto}
\end{figure*}

\begin{figure*}
    \centering
    \includegraphics[clip,trim=2.2cm 0.5cm 2.2cm 1cm, width=\textwidth,center]{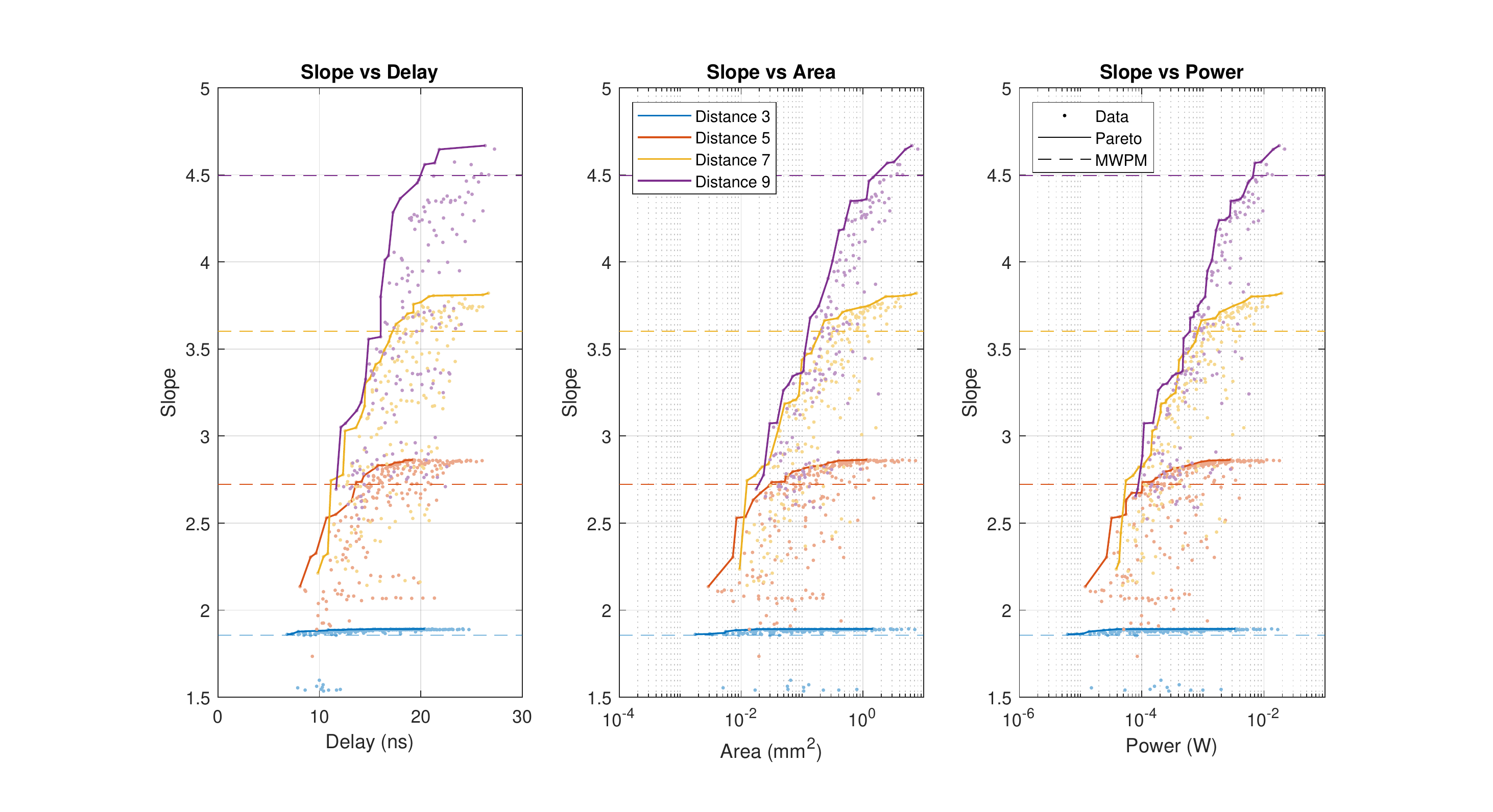}
    \caption{Slopes of the quantized neural network decoders versus their estimated hardware cost. The figure shows the Pareto front of the different distances and compares them to the MWPM decoder.}
    \label{fig:pareto_slope_all}
\end{figure*}

Figures \ref{fig:Pareto} and \ref{fig:pareto_slope_all} show the hardware cost estimates (delay, area, and power) of an ASIC design versus the obtained pseudo-threshold and slope, respectively.  For both figures the hardware cost is divided into three plots. The dots in these plots, colored according to the corresponding distance, represent all the quantized configurations as in Fig. \ref{fig:q_all_norm}. The solid lines show the Pareto front for every distance, i.e.,  the set of the neural networks that perform better for that particular trade-off. The horizontal dashed lines represent the performance of the  MWPM decoder. The hardware estimates of the individual nodes are included in \cite{DOI}.\\

As shown in  Fig. \ref{fig:Pareto}, a larger circuit is needed  to obtain the same $p_{th}$ at a larger distance. This trend holds until the Pareto fronts start to flatten out. The performance saturation is very clearly visible for distances 3 and 5. For distance 7 and 9, the curves also seem to saturate but this is only visible in the Pareto front and not in the whole data set. We would also expect $p_{th}$ to go on further, especially for distance 9, as larger distances are expected to have a larger $p_{th}$. The saturation of the Pareto front is therefore attributed to the finite size of our neural networks, which limits their performance at larger distances. Similar trends appear in Fig. \ref{fig:pareto_slope_all} on the slope data. Here the data points also show less saturation for the larger distances. However,  increasing the neural network sizes would be necessary  to prove this in future work.

The data on the decoder slope in Fig. \ref{fig:pareto_slope_all} also indicates a correlation between the hardware cost and the obtained performance. In contrast to the $p_{th}$, the Pareto fronts for the different distances seem to follow the same basic trend, and differentiate only by a different  saturation value.\\

The delays of the neural networks are all smaller than 30 ns, i.e., an order of magnitude lower than the required 440 ns, thus indicating that  less parallelism in the hardware implementation is feasible. A lower parallelism could help reducing both the area and power, which are quite large and scale exponentially with the performance. However, when using non-Clifford gates, a  decoding time faster than 440 is preferred  to correct the errors as soon as possible. Assuming the correction can be done by performing a single-qubit gate in  20 ns (see Table \ref{tab:sc_cycle_times}), ample margin is still available in time to be used by the decoder.

Not being limited by the delay,  there should be enough time to use recurrent neural networks (RNN). RNNs would increase the number of  inputs per node, as their own outputs in the previous cycle are concatenated with the outputs of the previous layer. It also opens the possibility to use the more complex cells of Long-Short Term Memory (LSTM) that  typically have longer delays. Finally, more (and thus perhaps smaller) layers could be used as required by CNNs, which  might need a deeper neural network, especially for larger distances.

Looking at the $p_{th}$ and slope plot together, two considerations can be drawn:
\begin{itemize}
    \item If the highest $p_{th}$ is preferred given certain hardware constraints, one can best choose a smaller surface code distance. As a high $p_{th}$ is mainly preferred when the physical error rates are higher, this means the qubits probably could not be integrated into larger distances anyway.
    \item If the highest slope is preferred, it does not make sense to choose a distance where the slope already starts to saturate. Thus, the largest distance is the best choice. Similar to the previous point, if a high slope is preferred, probably the qubits are performing quite well at a low physical error rate. This means that the  qubits can probably be used together to form larger logical qubits.
\end{itemize}

\begin{figure}
    \centering
    \includegraphics[clip,trim=0.5cm 0.5cm 0.5cm 0.2cm,width=0.55\textwidth,center]{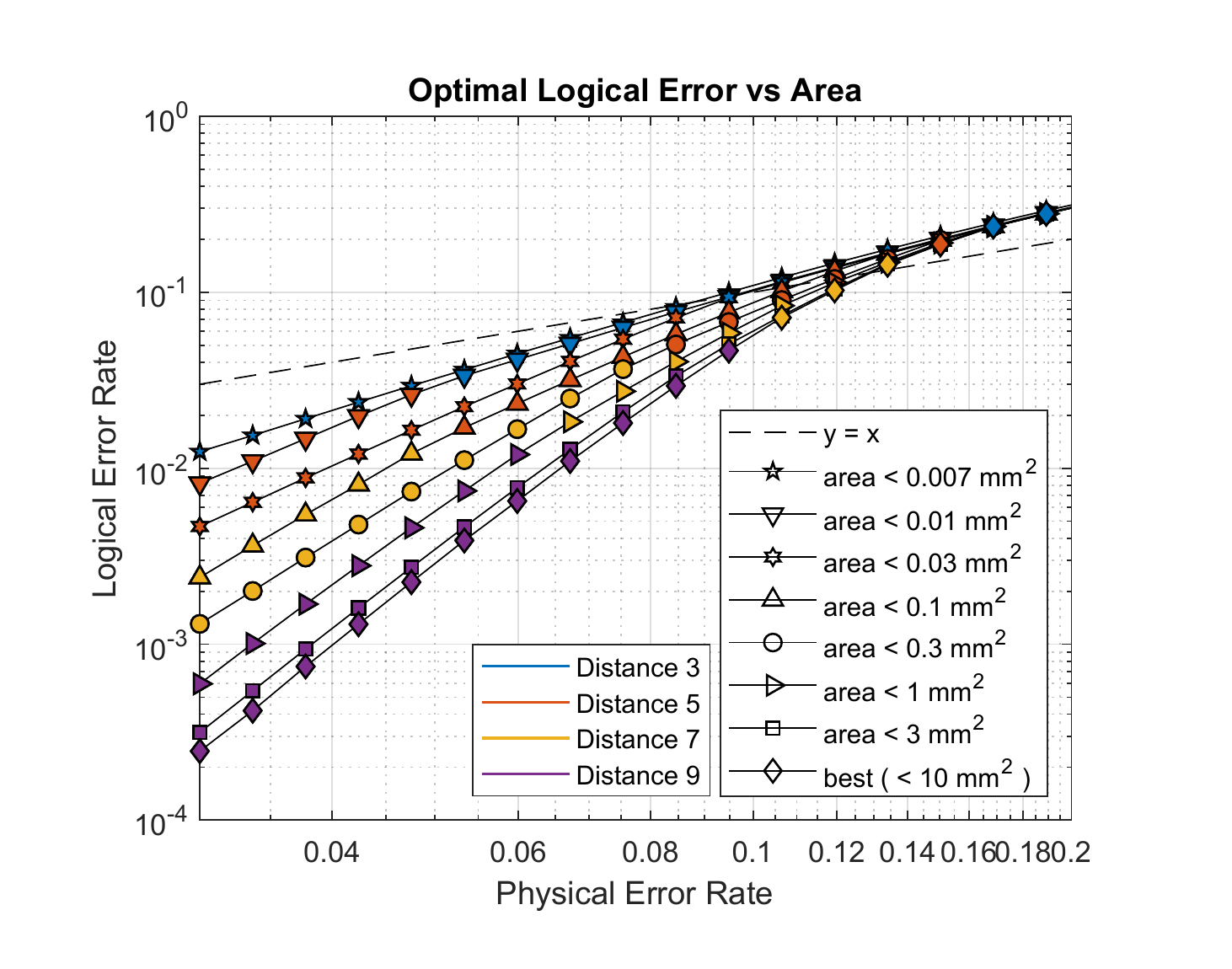}
    \caption{The optimal distance to obtain the lowest logical error rate for a certain physical error rate. The different lines indicate the maximum allowed area for the decoder.}
    \label{fig:logic_vs_areaf}
\end{figure}

Fig. \ref{fig:logic_vs_areaf} supports these conclusions. In this figure the optimum distance is shown for certain area constraints. Every line shows a different area constraint. This does not mean that this configuration will fully occupy this area but it will never exceed it. The markers on the lines change color depending on what distance obtains the lowest logical error rate for that physical error rate.
As discussed, lower distances perform better when the physical error rate is around the pseudo-threshold. In that region, as also Fig. \ref{fig:Pareto} illustrates, the best performance for a restricted area is always for lower distances. However, when the physical error rates become low enough, the slope starts to dominate. This is seen by a shift towards the larger distances for lower physical error rates.

\subsection{Comparing ASIC to FGPA}
We demonstrated that for near-term QEC, smaller distances are optimal. They obtain a higher $p_{th}$ given the same hardware. Near term QEC will also be a lot more experimental, requiring more frequent changes to the configuration of the neural network decoder and the corresponding weights and biases. 
As ASIC designs are  optimized for integration given a certain configuration, they might not be  suited for this task due to very limited reconfigurability. On the contrary, FPGAs can be reconfigured very easily and can even be synthesized to incorporate certain weights and biases, thus even  optimizing away unnecessary hardware. A downside to FPGAs is that their hardware is more generic and thus less optimized in terms of delay and power. Furthermore, their adoption imposes a strict limit in area, as only a fixed amount of hardware primitives (look-up tables, flip-flops) can be used.\\

To compare the FPGA to the ASIC designs, the same hardware was implemented on the Xilinx Artix-7 FPGA. However, instead of synthesizing all the nodes individually, the whole design was synthesized and implemented as a whole, thereby demonstrating if the design would actually fit on the selected  FPGA. The most promising designs were chosen, either the ones lying  on the Pareto front that would just barely obtain the MWPM performance or the ones for which the curves in figures \ref{fig:Pareto} and \ref{fig:pareto_slope_all} start to saturate.

After running synthesis and implementation using Xilinx Vivado, only the designs in Table \ref{tab:asicfpga} did actually fit on the FPGA. These are the distance-3 decoders achieving  maximum performance and performance comparable to MPWM and the distance-5 decoder with MPWM performance.All FPGA values are post-implementation room-temperature estimates. The power estimate is the reported dynamic power for a 440 ns clock period. The table shows that all decoders still fit the delay requirements by a large margin. This means that if this hardware would be optimized for area, e.g. by reducing the parallelism, probably larger neural networks could fit as well. 

Despite those limitations, the results are very promising, as even the FPGA designs can meet the required delay. The delay and hardware costs would be even lower if the synthesis was run with hard programmed weights but this was omitted for a fair comparison to the ASIC designs. 

\begin{table}
 \centering
 \caption{Three designs that fit on the Artix-7 FPGA and lie on the $p_{th}$ Pareto front. The table shows the distances and configuration, along with the decoding performance and hardware cost for both the FPGA and the ASIC design. All designs have $d^2-1$ inputs for layer 1 and two nodes in the output layer.}
\begin{tabular}{c r c c c}
\hline
        \hline
  \multicolumn{2}{r}{Distance} & 3 & 3 & 5 \\
  \hline\rule{0.0pt}{1\normalbaselineskip}
  Layer 1 & Size & 8 & 16 & 64 \\
  $\;$Layer 2 & Size & 4 & 4 & 64 \\
  \multicolumn{2}{r}{Bits}         & 3 & 5 & 4  \\
  \\
  \multicolumn{2}{r}{$p_{th}$} & 0.0823 & 0.0976 & 0.1037 \\
  \multicolumn{2}{r}{Slope} & 1.8641 & 1.8868 & 2.6641 \\
  \\
        & Delay & 17.9 ns & 71.9 ns & 87.6 ns \\
  FPGA  & Area  & 351 LUT & 2942 LUT & 44670 LUT \\  
        & Power & $<$ 1 mW  & 6 mW    & 132 mW\\
        \\
        & Delay & 7.3 ns & 12.3 ns & 14.3 ns \\
  ASIC  & Area & 0.0031 mm$^2$ & 0.0114 mm$^2$ & 0.3937 mm$^2$ \\
        & Power& 10.7 $\mu$W & 43.2 $\mu$W & 1.0 mW\\
        \hline
        \hline
    \end{tabular}
 \label{tab:asicfpga}
\end{table}

\subsection{Moving to cryogenic temperature}
The values reported for ASICs are extracted from simulations at 300 K. From these, we can draw some conclusions of the performance at 4.2 K.
The work of \cite[Fig.~10.1]{Homulle2019}\cite{CICCfabio} found that at cryogenic temperatures  the delay of digital cells decreases by  up to 50\% for  mature CMOS technologies thanks to the increase in mobility. However,the speed-up  will be much less significant in advanced commercial technologies as the increase in threshold voltage combined with the reduction of supply voltages mitigates those effect. As a result, the delay estimates  can be assumed to  approximately hold also at 4.2 K.
Due to the increased subthreshold slope \cite{tHart2018}, the leakage power at cryogenic temperatures is greatly reduced. Thus the power at cryogenic temperature is estimated to be lower than at 300 K.
Finally, due to the increase in mismatch \cite{tHart2018} and latch-up \cite{Deferm1990,Marshall2010}, a larger area might be needed at 4.2 K to decrease the mismatch and  to increase  the number of  well taps  to combat latch-up \cite{Schriek2018}.
\section{Discussion}\label{sec:Discussion} 
We have proposed  a new pure error decoder with more symmetries than previous works. These symmetries were also incorporated in the neural network resulting in improved performance. For future works, however, even more symmetries could be exploited, for example using a convolutional neural network. Another benefit of the novel pure error decoder is the equal delay of every chain.

A fully-connected feed-forward neural network with two hidden layers was used for the high-level decoder. Two hidden layers were chosen to minimize the delay and because this is enough to fit any possible function given enough nodes. The hardware estimates show, however, that the delay is small enough to allow for more layers.

For the space exploration, first the hyperbolic tangent transfer function was approximated by the SQNL function, which is significantly cheaper  in hardware cost and also outperforms the hyperbolic tangent.

Next, the layer sizes of the two hidden layers were swept and compared to the MWPM decoder and previous work \cite{Varsamopoulos2018_1, Varsamopoulos2018_2}. Even though our layers were significantly smaller, the obtained performance was on par with or better than previous research. The results also showed that, especially for distance 7 and 9, the performance could be further improved  by increasing the layer sizes or the neural network depth.
These results for a feed-forward neural network can be extended and compared to recurrent and convolutional neural networks in future work.

To reduce the hardware cost, all weights and outputs were quantized between 3 and 9 bits using a fractional fixed point two's complement representation.
For 9 bits, all distances performed on par with the floating point results before quantization.
Even fewer bits were needed to keep the performance above that of MWPM. More bits are needed  for larger distances, possibly  pointing to the need for some rescaling of the transferfunction domain depending on the layer size. Another solution might  be to use a sparsely connected or convolutional neural network. This solution could also exploit the translational symmetry of the surface code and our pure error decoder. The final option  to move towards less bits is to either use a binary neural network or to train using a different methodology than was used in this paper.

Finally, the ASIC hardware cost was estimated for every configuration in terms of delay, power and area. The data shows clear trends for both the pseudo-threshold and the decoding slope. The hardware cost for a certain pseudo-threshold is a lot higher for larger distances. On one hand, this means that, if the main objective is to obtain the lowest-cost hardware for physical qubits with a high error rate, the lowest distance that can obtain that pseudo-threshold should be chosen. On the other hand, if the physical qubits perform well below the pseudo-threshold, the data on the decoding slope clearly shows that there is a correlation between the steepest slope and the hardware costs. It also indicates that with an increasing distance, the maximum slope increases as well. If the qubits perform well enough and a large enough surface code can be made, it is desirable to choose the larger distance. 

The fully parallelized implementation chosen in this work  makes the needed hardware larger than necessary, as the 440 ns that are required to avoid a data backlog are met by a large margin with by the ASIC and the FPGA  designs. This would justify designs with  more hardware  reuse to trade-off the extra delay for smaller and lower-power solutions, which is needed to get the area well below $10 \mathrm{mm}^2$ and the power below $1$ W. Due to the reconfigurability of FPGAs, the weights could also be synthesized into the design, removing a lot of unnecessary hardware. Both these optimizations could significantly reduce the area and power. Combined with the exploration of CNNs,  this is a relevant direction for future research. Also the hardware cost versus performance of recurrent neural networks should be explored when using a more realistic error model including measurement errors, such as circuit noise. Employing recurrency and the circuit noise error model would allow the comparison with other competitive decoder algorithms, such as \cite{Das2021}. Furthermore, unlike other alternatives, the proposed  algorithm has the potential to outperform the MWPM algorithm. 


\section{Conclusion}\label{sec:Conclusion} 
This work presents an extensive space exploration of a high-level decoder for surface codes consisting of a pure error decoder and a fully-connected feed-forward neural network with two hidden layers. The results show that the decoder can be optimized for hardware simplicity, while still obtaining state-of-the-art decoding performance. The resulting hardware implementation allows the decoder to obtain decoding times less than 30 ns in both ASIC and FPGA implementations, which is significantly lower than the required 440 ns needed to keep up with the surface code cycles of current solid-state qubit technologies. The required area and power dissipation for the ASICs are realistic for a practical implementation at surface-code distances  up to 9 and for decoding performance well superior to the minimum-weight-perfect-matching algorithm. This  paves the way to quantum error correction hardware that can be co-integrated with the qubits at cryogenic temperatures.

\section*{Acknowledgments}
The authors would like to thank Savvas Varsamopoulos and Pinakin Padalia for their useful discussions. This work was supported by Intel.

\bibliographystyle{apsrev4-2}
\bibliography{bibmain}
\clearpage

\end{document}